%% file: main.tex
\documentclass[letterpaper, 10 pt, journal]{IEEEtran}
\pdfoutput=1
 \makeatletter

 \def\@testdef #1#2#3{%
   \def\reserved@a{#3}\expandafter \ifx \csname #1@#2\endcsname
  \reserved@a  \else
 \typeout{^^Jlabel #2 changed:^^J%
 \meaning\reserved@a^^J%
 \expandafter\meaning\csname #1@#2\endcsname^^J}%
 \@tempswatrue \fi}

\input{packages}

\input{macros}

\newtheorem{theorem}{Theorem}
\newtheorem{definition}{Definition}
\newtheorem{example}{Example}
\newtheorem{problem}{Problem}
\newtheorem{proposition}{Proposition}

\begin{document}
\title{Structured Synthesis for Probabilistic Systems}

\author{\IEEEauthorblockN{Nils Jansen\IEEEauthorrefmark{1}, Laura Humphrey\IEEEauthorrefmark{2}, Jana Tumova\IEEEauthorrefmark{3}, and Ufuk Topcu\IEEEauthorrefmark{1}}
\thanks{\IEEEauthorrefmark{1}Radboud University, Nijmegen, The Netherlands}%
\thanks{\IEEEauthorrefmark{2}Air Force Research Laboratory, USA}%
\thanks{\IEEEauthorrefmark{3}KTH Royal Institute of Technology, Sweden}%
\thanks{\IEEEauthorrefmark{1}University of Texas at Austin, USA}%
}

\maketitle
\thispagestyle{empty}
\pagestyle{empty}

\input{abstract}

\input{introduction}

\input{case_study}

\input{preliminaries}
\input{structured_model_repair}

\input{nonlinear}
\input{mdp_transformation}

\input{experiments}

\input{outlook}

\newpage
\bibliographystyle{plainyr}
\bibliography{literature}

\input{appendix}

\end{document}

%% file: packages.tex
\usepackage{amsmath,amssymb,amsfonts}
\usepackage{mathtools}
\usepackage[usenames,dvipsnames]{color}
\usepackage{xspace}
\usepackage{microtype}
\usepackage{todonotes}
\usepackage{colonequals}
\usepackage{wasysym}
\usepackage{breqn}
 \usepackage{booktabs}
 \usepackage{multirow}
\usepackage{multicol}

\usepackage{paralist}

\usepackage{subfigure}
\usepackage{url}
\usepackage{graphicx}
\usepackage{float}
\usepackage{algorithm}
\usepackage{listings}
\usepackage{nicefrac}
\usepackage{tikz} 
\usepackage{pgfplots}
\usepackage{wrapfig}

\lstdefinestyle{prismprogram}{
    belowcaptionskip=1\baselineskip,
    breaklines=true,
    showstringspaces=false,
    basicstyle=\footnotesize,
    keywordstyle=\normalfont\bfseries\color{black},
    identifierstyle=\itshape,
    morecomment=[s][\ttfamily]{[}{]}
}

\newlength{\MaxSizeOfLineNumbers}%
\usepackage{subfigure}
\usepackage{paralist}
\usepackage{booktabs}
\usepackage{parcolumns}

\def\createlinenumber#1#2{
    \edef\thelstnumber{%
        \unexpanded{%
            \ifnum#1=\value{lstnumber}\relax
              #2%
            \fi}%
        \ifx\thelstnumber\relax\else
        \expandafter\unexpanded\expandafter{\thelstnumber}%
        \fi
    }
}

\usetikzlibrary{arrows,decorations.pathmorphing,positioning,fit,trees,shapes,shadows,automata,calc} 

\tikzset{outline/.style args={#1}{%
  draw=#1,thick,fill=#1!50}}

%% file: macros.tex
\newcommand{\eqnref}[1]{(\ref{#1})}
\newcommand{\figref}[1]{Fig.~\ref{#1}}
\newcommand{\tabref}[1]{Table \ref{#1}}

\newcommand{\Distr}{\mathit{Distr}}

\newcommand{\distDom}{X}

\newcommand{\distFunc}{\mu}
\newcommand{\distDomElem}{x}

\newcommand{\biglor}{\ensuremath{\bigvee}}

\newcommand{\expr}{\mathit{Expr}}

\newcommand{\R}{\mathbb{R}}

\newcommand{\Z}{\mathbb{Z}}

\newcommand{\Ireal}{[0,1]\subseteq\mathbb{R}}  

\newcommand{\finally}{\lozenge}

\newcommand{\sinit}{s_{\mathit{I}}} 

\newcommand{\dtmc}{\mathcal{D}}

\newcommand{\mdp}{\mathcal{M}}
\newcommand{\MdpInit}[1][]{\ensuremath{\mdp{#1}=(S{#1},\sinit{#1},\Act,\probmdp{#1})}}
\newcommand{\pa}{\mdp}

\newcommand{\probmdp}{\mathcal{P}}

\newcommand{\act}[1][a]{\alpha} 
\newcommand{\Act}{\mathit{A}}        
\newcommand{\Sched}{\ensuremath{\mathit{Str}}}
\newcommand{\sched}{\ensuremath{\sigma}}

\newcommand{\p}{\ensuremath{\mathbb{P}}}
\newcommand{\pr}{\ensuremath{\mathrm{Pr}}}
\newcommand{\reachPr}[2]{\ensuremath{\pr^{#1}(\finally #2)}}
\newcommand{\reachPrT}[1][]{\ensuremath{\reachPr{#1}{T}}}
\newcommand{\reachProp}[2]{\ensuremath{\p_{\leq #1}(\finally #2)}}
\newcommand{\reachProplT}{\ensuremath{\reachProp{\lambda}{T}}}
\newcommand{\reachPrs}[3]{\ensuremath{\pr^{#1}_{#2}(\finally #3)}}
\newcommand{\reachPropSymbol}{\varphi}

\newcommand{\er}{\ensuremath{\mathrm{EC}}}
\newcommand{\ec}{\ensuremath{\mathrm{EC}}}
\newcommand{\expRew}[2]{\ensuremath{\er^{#1}(\finally #2)}}

\newcommand{\rewFunction}{\ensuremath{{\mathcal{C}}}}
\newcommand{\costFunction}{\ensuremath{\rewFunction}}

\newcommand{\ltp}[1][L]{\mathcal{L}}   

\newcommand{\actionof}[1]{\mathrm{act}(#1)}

\newcommand{\Vars}{\mathrm{Var}}

\newcommand{\PA}{\pa}

\newcommand{\true}{\ensuremath{\texttt{true}}\xspace}

\DeclareMathOperator{\dom}{dom}
\newcommand{\ie}{i.\,e.\xspace}
\newcommand{\eg}{e.\,g.\xspace}

\newcommand{\var}{\ensuremath{x}}

\newcommand{\bddstate}[1]{s_{\mathbb{B}}{}}

\newcommand{\Var}{\ensuremath{V}}
\newcommand{\Valuations}{\ensuremath{\mathit{Val}}}

\newcommand{\Ass}{\mathcal{V}}
\newcommand{\Assignments}{\mathcal{V}}                 

\newcommand{\tool}[1]{\texttt{#1}\xspace}
\newcommand{\prophesy}{\tool{PROPhESY}}

\newcommand{\prism}{\tool{PRISM}}
\DeclareRobustCommand{\cpp}
{\valign{\vfil\hbox{##}\vfil\cr
   \textsf{C\kern-.1em}\cr
   $\hbox{\fontsize{\sf@size}{0}\textbf{+\kern-0.05em+}}$\cr}%
}

\selectcolormodel{cmy}

\definecolor{lgrey}{rgb}{0.8,0.8,0.8}
\definecolor{grey}{rgb}{0.5,0.5,0.5}

\definecolor{lightblue}{rgb}{0.8,0.8,1.0}
\definecolor{lightred}{rgb}{1.0,0.8,0.8}
\definecolor{lightgreen}{rgb}{0.8,1.0,0.8}
\definecolor{angrygreen}{cmyk}{0.279,0,0.91,0.08}
\definecolor{lightred}{rgb}{1.0,0.8,0.8}
\definecolor{pink}{rgb}{1.0,0.1,1.0}

\newcommand{\nj}[1]{\todo[inline]{\scriptsize{NJ: #1}}}

\definecolor{prismgreen}{HTML}{009900}
\definecolor{prismred}{HTML}{cc0000}
\definecolor{prismblue}{HTML}{0000cc}

\lstdefinelanguage{Prism}{
        basicstyle=\color{prismred}\scriptsize\ttfamily,
        literate=*	{0}{{\textcolor{prismblue}{0}}}{1}
			{1}{{\textcolor{prismblue}{1}}}{1}
			{2}{{\textcolor{prismblue}{2}}}{1}
			{3}{{\textcolor{prismblue}{3}}}{1}
			{4}{{\textcolor{prismblue}{4}}}{1}
			{5}{{\textcolor{prismblue}{5}}}{1}
			{6}{{\textcolor{prismblue}{6}}}{1}
			{7}{{\textcolor{prismblue}{7}}}{1}
			{8}{{\textcolor{prismblue}{8}}}{1}
			{9}{{\textcolor{prismblue}{9}}}{1}
			{.0}{{\textcolor{prismblue}{.0}}}{2}
			{.1}{{\textcolor{prismblue}{.1}}}{2}
			{.2}{{\textcolor{prismblue}{.2}}}{2}
			{.3}{{\textcolor{prismblue}{.3}}}{2}
			{.4}{{\textcolor{prismblue}{.4}}}{2}
			{.5}{{\textcolor{prismblue}{.5}}}{2}
			{.6}{{\textcolor{prismblue}{.6}}}{2}
			{.7}{{\textcolor{prismblue}{.7}}}{2}
			{.8}{{\textcolor{prismblue}{.8}}}{2}
			{.9}{{\textcolor{prismblue}{.9}}}{2}
			{[}{{\textcolor{black}{[}}}{1}
			{]}{{\textcolor{black}{]}}}{1}
			{+}{{\textcolor{black}{+}}}{1}
			{-}{{\textcolor{black}{-}}}{1}
			{=}{{\textcolor{black}{=}}}{1}
			{<}{{\textcolor{black}{<}}}{1}
			{>}{{\textcolor{black}{>}}}{1}
			{\&}{{\textcolor{black}{\&}}}{1}
			{|}{{\textcolor{black}{|}}}{1}
			{:}{{\textcolor{black}{:}}}{1}
			{;}{{\textcolor{black}{;}}}{1}
			{(}{{\textcolor{black}{(}}}{1}
			{)}{{\textcolor{black}{)}}}{1}
			{..}{{\textcolor{black}{..}}}{2},
        keywords= {bool,C,ceil,const,ctmc,double,dtmc,endinit,endmodule,endrewards, endsystem,F,false,floor,formula,G,global,I,init,int,label,max,mdp,min,module,nondeterministic,P,Pmin,Pmax,prob,probabilistic,R,rate,rewards,Rmin,Rmax,S,stochastic,system,true,U,X},
        keywordstyle={\bfseries\color{black}},
        numberstyle=\footnotesize\color{black},
        comment=[l] {//}, morecomment=[s]{/*}{*/},
        commentstyle= \color{prismgreen},
        tabsize=4,
        captionpos=b,
        escapechar=@
}


%% file: abstract.tex
\begin{abstract}
We introduce the concept of structured synthesis for Markov decision processes where the structure is induced from finitely many pre-specified options for a system configuration.
The resulting synthesis problem is in general a nonlinear programming problem (NLP) with integer variables.
As solving NLPs is in general not feasible, we present an alternative approach.
We present a transformation of models specified in the {PRISM} probabilistic programming language to models that account for all possible system configurations by means of nondeterministic choices.
Together with a control module that ensures consistent configurations throughout the system, this transformation enables the use of optimized tools for model checking in a black-box fashion.
%
While this transformation increases the size of a model, experiments with standard benchmarks show that the method provides a feasible approach for structured synthesis.
Moreover, we demonstrate the usefulness along a realistic case study involving surveillance by unmanned aerial vehicles in a shipping facility.
\end{abstract}

%% file: introduction.tex
\section{Introduction}
\noindent The main problem introduced in this paper is motivated by the following scenario stemming from the area of physical security. 
Consider a shipping facility equipped with a number of ground sensors to discover potential intruders. 
The facility operates unmanned aerial vehicles (UAVs) to perform surveillance and maintenance tasks.
Intruders appear randomly, there are uncertainties in sensor performance, and the operation of the UAVs is driven by scheduling choices and the activation of sensors.
The natural model to capture such randomization, uncertainty, and scheduling choices are 
\emph{Markov decision processes} (MDPs), where
measures such as ``the probability to encounter dangerous states of the system'' or performance measures such as ``the expected cost to achieve a certain goal'' are directly assessable.
%

System designers, \eg in the shipping facility scenario, often have to choose among a pre-specified family of possibly interdependent options for the system configuration, such as different sensors or the operating altitude of UAVs.
Each of these options triggers different behavior of the system, such as different failure probabilities and acquisition cost. 
%
We call such possible design choices and their behavioral consequences an underlying \emph{structure of the system}; all concrete instantiations of the system adhere to this structure.
For instance, imagine a structure describing the option of installing one of two types of sensors. The cheaper sensor induces a smaller expected cost, while the more expensive sensor induces a higher probability of discovering intruders. 
The changes in the instantiations of the system are necessarily according to the structure, \ie, the replacement of one sensor type with the other. 
A question of interest is then which instantiation yields the lowest cost while it guarantees to adhere to a target specification regarding the desired probability.



In this paper, we introduce 
\emph{multiple--instance MDPs} (MIMDPs) as an underlying semantic model for structured synthesis.
Arbitrary expressions over system parameters capture a structure that describes dependencies between uncertain behavior and system cost.
Each parameter may be assigned a discrete finite set of values out of which it can be instantiated. 
MIMDPs are inspired by 
\emph{parametric Markov decision processes}~\cite{DBLP:journals/ior/SatiaL73,DBLP:journals/ai/DelgadoBDS16} (pMDPs), whose transition probabilities (and cost) are defined by functions over yet-to-be-specified parameters. 
However, the existing definitions of pMDPs are limited in that they only allow for imposing restrictions on parameter valuations in the form of continuous intervals. 
Available methods as implemented in the tools \tool{PARAM}~\cite{param_sttt}, \tool{PRISM}~\cite{KNP11}, or \tool{PROPhESY}~\cite{dehnert-et-al-cav-2015} do not support the definition of discrete sets of valuations.
Consequently, to the best of our knowledge, the state-of-the-art techniques cannot directly handle modeling scenarios that appear in structured synthesis. 

Another related approach to modeling structured systems is \emph{feature-based modeling} as in \tool{ProFeat}~\cite{profeat}, which allows to specify families of stochastic systems that have the same base functionality, but differ in the combination of active and inactive features that trigger additional functionalities. 
Although feature-based modeling supports discrete parametrization, it does not directly offer parametrization of probabilities.
Furthermore, \tool{ProFeat} enables the analysis of the family in an all-in-one fashion as opposed to focusing on properties of individual instantiations.

The formal problem considered in this paper is to compute an optimal instantiation of parameters and a control strategy for a given MIMDP subject to reachability specifications and cost minimization. 
We define this problem naturally  as a \emph{non--linear integer  optimization problem} (NILP).

As a computationally tractable alternative, we present a transformation of an MIMDP to an MDP where all possible parameter instantiations are relayed to \emph{nondeterministic choices}.
The common language used for model specification in all available tools is the probabilistic programming language~\cite{DBLP:conf/icse/GordonHNR14} originally developed for \tool{PRISM}.
We define the transformation of the MIMDP as a \emph{program transformation} of such a probabilistic program.
%
By adding control variables to the transformed program, we keep track of all instantiations, ensuring parameter values that are in accordance with the given structure.
We show that optimal measures for the resulting MDP are optimal for the NILP defining our original problem.
Computing an optimal solution to the original problem is thereby reduced to \emph{MDP model checking}, which is equivalent to solving a linear program (LP) and hence computationally more tractable.
From a practical viewpoint, the transformation enables the use of all capabilities of model checkers such as \tool{PRISM}~\cite{KNP11}, \tool{storm}~\cite{storm}, or \tool{IscasMC}~\cite{iscasmc}.

We illustrate the feasibility of the proposed approach on several basic case studies from the \prism benchmark suite~\cite{KNP12b}.
We also report promising results for a more realistic case study based on the shipping facility example. 
In our experiments, we observe that the transformation from a MIMDP to an MDP involves an increase in the number of states or transitions of up to two orders of magnitude. 
However, using an efficient model checker,
we are able to demonstrate the applicability of our approach in examples with millions of states.

In summary, the contribution of this paper is threefold: i) We define a parametric model supporting discrete valuation sets and formalize the structured synthesis problem.
ii) We develop a transformation of the parametric model to the \tool{PRISM} language allowing us to practically address the structured synthesis problem. iii) We present a detailed, realistic case study of a shipping facility as well as experimental evaluation on standard benchmarks.

%% file: case_study.tex
\section{Case Study}\label{ex:shipyard}\label{sec:casestudy}
\begin{figure}[t]
      \centering
      \scalebox{0.32}{\includegraphics[]{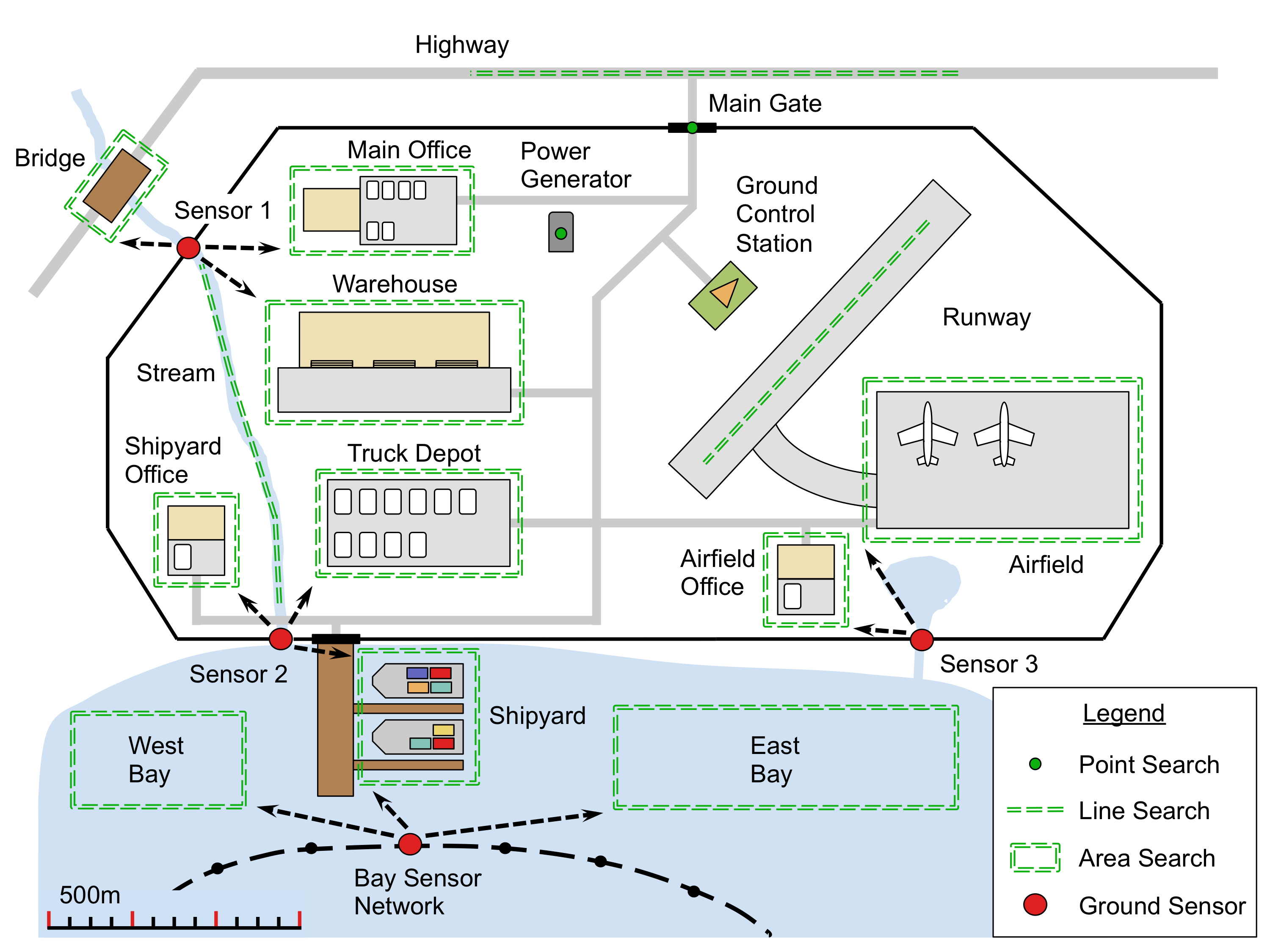}}
      \caption{A shipping facility that uses UAVs to perform surveillance tasks.}
      \label{fig:scenario}
\end{figure}
In this section we introduce the case study that originally motivated the problem and the proposed approach. Several technical details are in the Appendix.

\paragraph*{Scenario}
Consider a hypothetical scenario in which a shipping facility uses one or more UAVs equipped with 
electro-optical (EO) sensors to perform surveillance tasks over various facility assets as shown in \figref{fig:scenario}. 
These assets include an \emph{Airfield} with a \emph{Runway} and \emph{Airfield Office}, 
a \emph{Truck Depot}, a \emph{Warehouse}, a \emph{Main Office}, a \emph{Shipyard} with a \emph{Shipyard Office}, 
and a small bay that is partitioned into \emph{West Bay}, \emph{East Bay}, and \emph{Shipyard} areas. 
An external \emph{Highway} connects to the \emph{Main Gate},  
and a nearby \emph{Bridge} crosses a \emph{Stream} that cuts through the facility and empties into the bay. 
The facility is surrounded by a fence, but points where 
waterways run under the fence might allow human intruders to enter. 
These points and the bay are monitored by \emph{Sensor 1}, \emph{Sensor 2}, \emph{Sensor 3}, and a \emph{Bay Sensor Network}.
All of these \emph{ground sensors} can detect intruders with an adjustable false alarm rate 
and record the direction in which detected intruders were moving.

UAVs take off and land from an area near the \emph{Ground Control Station} (GCS). 
For each of the facility assets, a UAV can be tasked to perform 
a point, line, or area search as indicated in \figref{fig:scenario}. 
Each of these search tasks requires the UAV to fly a certain distance to get to the task location, 
carry out the task, and fly back to the GCS. 
For point searches, simply flying to the point and back is enough. 
For line searches, the UAV must fly to one end, follow the line, and fly back from the other end. 
For area searches, the UAV must fly to one corner of the area, perform sequential parallel passes over it 
until the entire area has been covered by the UAV's sensor footprint, then fly back from the terminating corner. 
Note that sensor footprint size decreases with altitude, so more passes are needed to cover an area as a UAV's altitude decreases.

A UAV can also be tasked to fly to and query the ground sensors. 
If a ground sensor reports that an intruder has been detected since the last time a UAV visited, 
the UAV flies to the area the intruder was heading toward and performs a search;
otherwise, the UAV can return to the GCS and continue on to another task. 
Probabilities for which areas intruders are likely to head toward might be estimated over time or assumed to be uniform if data are not available. 
We assume surveillance occurs frequently enough that at most one intruder will pass by a ground sensor before it is queried by a UAV. 
Similar problems involving UAVs searching for intruders based on ground sensor information are discussed in, 
e.g., \cite{casbeer_2014} and \cite{kingston_2015}.

\paragraph*{System Configuration, Safety, and Cost} A configuration of the shipyard facility refers to the types of ground sensors that are used and the types of EO sensors installed onboard the UAVs. 
We measure \emph{safety} of the shipping facility in terms of the probability to successfully detect intruders. 
The \emph{performance} measure describes the expected cost for the shipping facility given the system uncertainties. 
Different types of sensors provide different tradeoffs between safety and performance. 
Our goal is to find a configuration for the shipyard surveillance system that ensures 
a certain safety probability on detecting intruders while minimizing cost. 

Different ground sensor types and EO sensor types result in safety and performance tradeoffs for several reasons. 
First, each sensor type has a different one-time purchase cost. 
In turn, each sensor type has tunable parameters that result in 
a tradeoff between the probability of detecting an intruder and cost in terms of UAV flight time, 
with sensors that have a higher purchase cost providing a better tradeoff. 
The tradeoff between intruder detection and UAV flight time is also affected by UAV operating altitude, 
which is adjustable.

This tradeoff can be understood in the context of two factors. 
The first is \emph{ground sample distance} (GSD)~\cite{kingston2016automated}, 
\ie, the number of meters per pixel of images sent back by a UAV, which depends on UAV altitude and EO sensor resolution.
The second is the ground sensor \emph{receiver operating characteristic}~(ROC) \cite{fawcett_2006}, 
\ie, the tunable true positive versus false positive rate. 
Both true and false positives result in a UAV performing an area search for intruders, 
where the cost of the area search depends on UAV altitude.

We now describe in detail how the probabilistic parameters relating to GSD and ROC can be adjusted by acquiring different types of sensors, tuning sensor parameters, or changing UAV operating altitude, 
and how these choices affect cost.



\paragraph*{Basic Task Costs}
\label{sec:basicTaskCosts}

For tasks that do not involve intruder detection, 
cost is driven mainly by manpower, logistics, and maintenance requirements, 
which roughly corresponds to cost per 
flight second $c_f$. 
Suppose the UAVs in this scenario all fly at some standard operating ground speed $v_g$ measured in meters per second. 
The cost $c(t)$ for a task $t$ that does not involve intruder detection then depends in a straightforward way on the distance $d(t)$ that a UAV must fly:
\begin{equation}
c(t) = d(t) c_f/v_g.
\label{e:basicCost}
\end{equation}

%

\paragraph*{Image GSD}
\label{s:ImageGsd}

An important consideration for UAV surveillance tasks is the amount of visual detail a human operator needs to perform analysis of collected images, especially for small objects such as human intruders. 
The amount of resolvable detail depends on the number of pixels comprising objects of interest in the images, 
which increases as GSD decreases. 
%
GSD can be decreased by either \emph{decreasing altitude} or \emph{increasing horizontal resolution} of the EO sensor. 
On the other hand, altitude can be changed during flight, sensor resolution cannot, since it depends on the specifications of the installed EO sensor. For this scenario, we consider three common EO sensor resolution options (480p, 720p, 1080p), with hypothetical purchase prices (\$15k, \$30k, \$45k, respectively). 

%

We can use GSD in conjunction with the Johnson criteria~\cite{johnson_1958} to estimate the probability that a human operator can successfully analyze an object in an image based on the type of analysis task (i.e., \emph{detection}, \emph{recognition}, or \emph{identification}) and the number of pixels comprising an object of interest in an image.
For each type of task and a corresponding digital image, a quantity $n_{50}$ defines the number of pairs of pixel lines across the ``critical'' or smaller dimension of an object needed for a $50\%$ probability of task success.
For instance, $n_{50} =1$ for object detection.

Given $n_{50}$ and the actual number of pixels pairs $n$ across the critical dimension of an object (which depends on the size of the object and GSD), 
the probability for task analysis success $p_d$ given sufficient time to analyze the image is estimated as
\begin{equation}
p_d = \frac{(n/n_{50})^{x_0}}{1+(n/n_{50})^{x_0}} \quad\textrm{where}\quad x_0 = 2.7 + 0.7(n/n_{50}).
\label{eq:pJohnson}
\end{equation}
%

In our approach, we use a quadratic approximation of $p_d$ for simplicity. 
Since \eqref{eq:pJohnson} varies significantly with altitude,
we define a different operating altitude for each type of EO sensor in Appendix~\ref{s:altitude}.

\begin{figure}[thpb]
      \centering
      \includegraphics[width=.6\columnwidth]{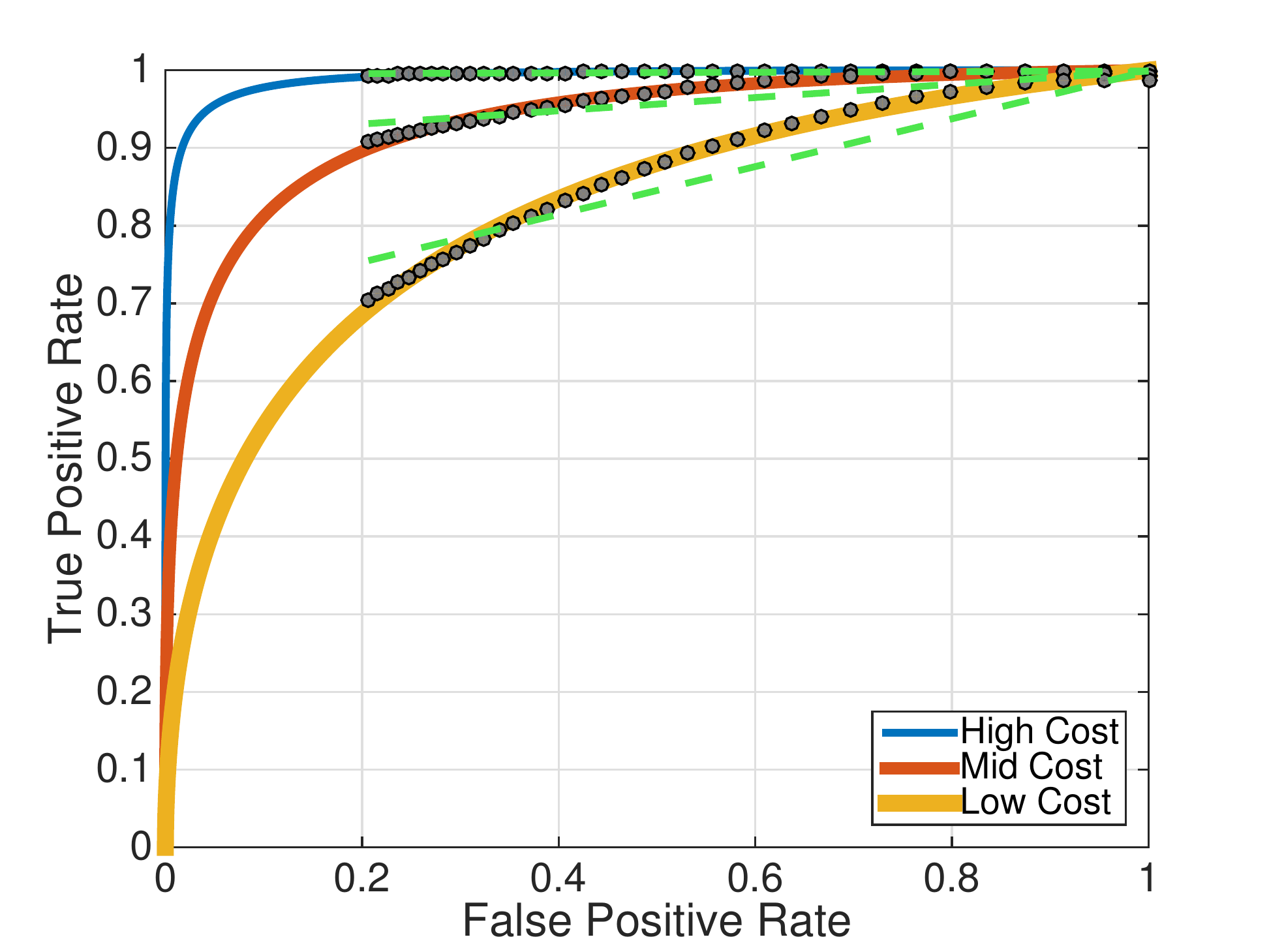}
      \caption{ROC curves for different cost ground sensors. Linear and quadratic approximations are shown as green dashed lines and traces of black dots, respectively.}
      \label{fig:rocCurvesApprox}
\end{figure}

\paragraph*{Ground Sensor ROC}
\label{s:groundSensorRoc}
The ROC curve of a sensor performing binary classification describes the tradeoff between the sensor's true positive rate/probability versus false positive rate/probability as the sensor's discrimination threshold is varied. 
Consider the three solid curves in \figref{fig:rocCurvesApprox}. 
These represent hypothetical ``low'', ``mid'', and ``high'' cost ground sensors, with one-time purchase costs of \$15k, \$30k, and \$45k, respectively. 
For each such ground sensor, the discrimination threshold can be varied to achieve an operating point on the corresponding curve. 
As the curves show, a high cost ground sensor provides the best tradeoff, since for each false positive rate, it provides a higher true positive rate. 
%
%

In our approach, we use the quadratic approximations of the curves in \figref{fig:rocCurvesApprox} over the interval $[0.2, 1.0]$ for false positive rates. In order to reliably detect intruders, we need the true positive rate to be fairly high. 
%

To help understand the effect ground sensors have on system costs and probabilistic parameters, suppose we choose to purchase a high cost sensor for Sensor 1. 
Clearly the purchase cost is higher than if we had chosen a mid or low cost sensor. 
However, each time the ground sensor generates a false positive, a UAV has to perform an unnecessary area search, which incurs additional system operational cost.
To counteract this, we could decrease the ground sensor's false positive rate, but this would also decrease its true positive rate, resulting in a higher false negative rate. 
When a false negative occurs, the sensor fails to detect the presence of an intruder, the UAV does not perform an area search, and the intruder goes undetected, potentially causing the system to violate its specification on the probability of detecting intruders.
A high cost sensor then mitigates operational cost by providing a lower false positive rate.  
Given these tradeoffs between purchase cost, operational cost, and probability of intruder detection, it is not clear which sensor minimizes cost while meeting safety specifications on intruder detection.

%
In summary, each configuration induces a different system behavior in terms of 
probabilities for discovery of intruders and task cost. 
The question is, which configuration ensures satisfaction of a given specification while minimizing the overall cost.

%% file: preliminaries.tex
\section{Preliminaries}\label{sec:preliminaries}
%
A \emph{probability distribution} over a finite or countably infinite set $\distDom$ is a function $\distFunc\colon\distDom\rightarrow\Ireal$ with $\sum_{\distDomElem\in\distDom}\distFunc(\distDomElem)=\distFunc(\distDom)=1$. 
The set of all distributions on $\distDom$ is $\Distr(\distDom)$.

\begin{definition}{\bf (MDP)}
A \emph{Markov decision process (MDP)} $\MdpInit$ consists of a finite set of states $S$, a unique initial state $\sinit \in S$, a finite set $\Act$ of actions, and a probabilistic transition function $\probmdp\colon S\times\Act\times S\rightarrow\Ireal$ with $\sum_{s'\in S}\probmdp(s,\act,s')=1$ for all $s\in S,\act\in\Act$.
\end{definition}
MDPs operate by means of \emph{nondeterministic choices} of actions at each state, whose successors are then determined \emph{probabilistically} with respect to the associated probabilities.
The \emph{enabled} actions at state $s\in S$ are denoted by $\Act(s)=\{\act\in\Act\mid\exists s'\in S.\,\probmdp(s,\act,s')>0\}$. 
To avoid deadlock states, we assume that $|\Act(s)|\geq 1$ for all $s\in S$.
A \emph{cost function} $\costFunction \colon S\rightarrow\R_{\geq 0}$ for an MDP $\mdp$ adds cost to a \emph{state}. 
%
%
If $|\Act(s)|=1$ for all $s\in S$, all actions can be disregarded and the MDP $\mdp$ reduces to a \emph{discrete-time Markov chain (MC)}.
%
In order to define a probability measure and expected cost on MDPs, the nondeterministic choices of actions are resolved by so-called \emph{strategies}.
For practical reasons, we restrict ourselves to \emph{memoryless} strategies, refer to~\cite{BK08} for details.
\begin{definition}{\bf (Strategy)}\label{def:scheduler}
	A \emph{randomized strategy} for an MDP $\mdp$ is a function $\sched\colon S\rightarrow\Distr(\Act)$ such that $\sigma(s)(a) > 0$ implies $a \in \Act(s)$. A strategy with $\sched(s)(a)=1$ for $a\in\Act$ and $\sched(b)=0$ for all $b\in\Act\setminus\{a\}$ is called \emph{deterministic}. The set of all strategies over $\mdp$ is denoted by $\Sched^\mdp$.
\end{definition}
%
Resolving all nondeterminism for an MDP $\mdp$ with a strategy $\sched\in\Sched^\mdp$ yields an \emph{induced Markov chain} $\mdp^\sched$. 

\begin{definition}{\bf (Induced MC)}\label{def:induced_dtmc}
	Let MDP $\MdpInit$ and strategy $\sched\in\Sched^{\mdp}$. The \emph{MC induced by $\mdp$ and $\sched$} is  $\mdp^\sched=(S,\sinit,\Act,\probmdp^\sched)$ where
	\begin{align*}
		\probmdp^\sched(s,s')=\sum_{a\in\Act(s)} \sched(s)(a)\cdot\probmdp(s,a,s') \quad \mbox{ for all } s,s'\in S~.
	\end{align*} 
\end{definition}
\paragraph{\tool{PRISM}'s Guarded Command Language.}
\label{subsec:gcl}
We briefly introduce the syntax and semantics of the probabilistic programming language used to specify probabilistic models in \tool{PRISM}.
For a finite set $\Vars$ of integer variables, let $\Ass(\Vars)$ denote
the set of all variable valuations, \ie of functions $u\colon\Vars\to\Z$
such that $u(x)\in\dom(x)$ for all $x\in\Vars$. We assume that the domains
of all variables are finite.
\begin{definition}[Probabilistic program, module, command]
  \label{def:prism_model}
  A \emph{probabilistic program} $(\Vars,\sinit,\{M_1,\ldots,M_k\})$ consists of a set of integer variables
  $\Vars$, an initial variable valuation
  $\sinit \in \Assignments(\Vars)$ and a finite set of modules $\{M_1,\ldots,M_k\}$.
  A \emph{module} is a tuple $M_i = (\Vars_i,\Act_i,C_i)$ with
  $\Vars_i\subseteq\Vars$ a set of variables such that
  $\Vars_i\cap\Vars_j=\emptyset$ for $i\not= j$, $\Act_i$ a finite set
  of synchronizing actions, and $C_i$ a finite set of commands. The
  action $\tau$ with $\tau\not\in\bigcup_{i=1}^k\Act_i$ denotes the
  internal non-synchronizing action. A \emph{command} $c\in C_i$ has the form\normalfont
  \begin{lstlisting}
	[$\act$] $g \rightarrow  p_1\colon f_1 + \ldots + p_n\colon f_n;$	
\end{lstlisting}\itshape
  with $\act\in\Act_i\uplus\{\tau\}$, $g$ a Boolean predicate
    (``guard'') over the variables in $\Vars$, $p_j\in\Ireal$ with $\sum_{j=1}^n p_j = 1$, and $f_j\colon
    \Assignments(\Vars)\to\Assignments(\Vars_i)$ being a variable
    update function. The action $\act$ of command $c$ is
    $\actionof{c}$.
\end{definition}
%
A model with several modules is equivalent to a model with a single
module, obtained by computing the \emph{parallel composition} of
these modules. For details we refer to the documentation of \tool{PRISM}.

\subsubsection{Specifications.}
We consider \emph{reachability properties} and \emph{expected cost properties}.
For MC $\dtmc$ with states $S$, let $\reachPrs{\dtmc}{s}{T}$ denote the probability of reaching a set of \emph{target states} $T \subseteq S$ from  state $s\in S$; simply $\reachPrT[\dtmc]$ denotes the probability from initial state $\sinit$.
For threshold $\lambda\in [0,1]$, a \emph{reachability property} asserts that a target state is to be reached with probability at most $\lambda$, denoted by $\reachPropSymbol = \reachProplT$.
The property is satisfied by $\dtmc$, written $\dtmc \models \reachPropSymbol$, iff $\reachPrT[\dtmc]\leq\lambda$.

The cost of a path through MC $\dtmc$ to a set of \emph{goal states} $G\subseteq S$ is the sum of action costs visited along the path. The expected cost of a finite path is the product of its probability and its reward.
For $\reachPr{\dtmc}{G} = 1$, the expected cost $\expRew{\dtmc}{G}$ of reaching $G$ is the sum of expected costs of all paths leading to $G$. 
Using recent results from~\cite{arnd_tacas}, we also consider the probability $\pr^{\dtmc}(\finally T \land C<n)$, where the total cost $C$, \ie, the sum of the costs of all paths satisfying $\finally T$, is bounded by $n$.
%
Formal definitions are given in e.g.,~\cite{BK08}.

%% file: structured_model_repair.tex
\section{Structured Synthesis}\label{sec:structured}
%
In this section, we introduce the formal model to define a structure on MDPs, and we define the synthesis problem.


%
\subsection{Multiple-instance MDPs}

We use a variant of a parametric Markov decision process (pMDPs) which we call a \emph{multiple--instance Markov decision process} (MIMDP).
Formally, we define a finite set $\Var=\{p_1,\ldots,p_n\}$ of parameters. 
Each parameter $p\in\Var$ has a finite range of values $\Valuations(p)=\{v_1,\ldots,v_m\} \subseteq \mathbb R$.
A \emph{valuation} is a function $u \colon\Var \rightarrow \bigcup_{p\in \Var}{\Valuations(p)}$ that respects the parameter ranges, meaning that for a parameter $p$, $u(p) = v \in \Valuations(p)$. 
Let $U(\Var)$ denote the (finite) set of possible valuations on $\Var$. 

Let $\expr(\Var)$ denote the set of \emph{expressions} over $\Var$ and 
$p\in l$ state that parameter $p$ occurs in expression $l\in\expr(\Var)$. $\Valuations(l)$
denotes the (finite) set of possible values for $l\in\expr(\Var)$ according to the parameters $p\in l$ and their value ranges $\Valuations(p)$.
With a slight abuse of notation, we lift valuation functions from $U(V)$ to expressions: $u \colon \expr(\Var) \rightarrow \bigcup_{l \in \expr(\Var)}{\Valuations(l)}$.
In particular, $u(l) = v \in \Valuations(l)$ is the valuation of $l$ obtained by the instantiation of each $p \in l$ with $u(p)$.
Note that for a particular valuation of two or more expressions $l, l'$ in $\expr(\Var)$, there is nothing constraining the valuation of the underlying parameters to be consistent. That is, a parameter valuation $u(p)$ that results in expression valuation $u(l)$ might be different than a $u'(p)$ that results in $u(l')$.
%
%
\begin{example}
Consider $\Var = \{p,q\}$, $\Valuations(p) = \{0.1,0.2\}$, $\Valuations(q) = \{0.3,0.4\}$, and $\expr(\Var) = \{p+q, p + 2\cdot q\}$. The ranges of values for the expressions are $\Valuations(p+q) = \{0.4, 0.5, 0.6\}$, and $\Valuations(p + 2 \cdot q) = \{0.7, 0.8, 0.9, 1\}$. A possible valuation on the parameters is $u(p) = 0.1, u(q) = 0.3$ determining an associated valuation on the expressions $ u(p+q) = 0.4,  u(p + 2\cdot q) = 0.7$.
 A possible valuation on the expressions is $\upsilon(p+q) = 0.4$, $\upsilon(p+ 2\cdot q) = 1$, however, this valuation is not consistent with any valuation on the parameters; the first is consistent with $u(p)=0.1$ and $u(q)=0.3$, while the second is consistent with $u(p)=0.2$ and $u(q)=0.4$.
\end{example}
\begin{definition}{\bf (MIMDP)}
A \emph{multiple-instance MDP} $\mdp=(S,\Var,\sinit,\Act,\probmdp)$ has a finite set $S$ of states, a finite set $\Var$ of parameters with associated finite sets of valuations from $\Valuations(\Var)$, a unique initial state $\sinit \in S$, a finite set $\Act$ of actions, and a transition function $\probmdp\colon S\times\Act\times S\rightarrow \expr(\Var)$. 
\end{definition}
Similarly to MDPs, we use $A(s) \subseteq A = \{\alpha \mid P(s,a,s') \text{ is defined for some } s'\in S\}$ to denote the set of actions enabled in state $s \in S$.
A \emph{cost function} $\costFunction\colon S\rightarrow \expr(\Var)$ associates (parametric) cost to states.
For each valuation $u\in\Valuations(\Var)$ of parameters, the \emph{instantiated MIMDP} is $\mdp[u]$. 
We denote the set of all expressions occurring in the MIMDP by $\mathcal{L}_\mdp$.

Note that an MIMDP is a special kind of parametric MDP (pMDP)~\cite{param_sttt,quatmann-et-al-atva-2016} in the sense that each parametric cost and probabilities can only take a finite number of values, \ie, there are multiple but finite instantiations of a MIMDP.
The state--of--the--art tools such as \tool{PARAM}~\cite{param_sttt}, \tool{PRISM}~\cite{KNP11}, or \prophesy~\cite{dehnert-et-al-cav-2015}, however, only allow for defining continuous intervals to restrict parameter valuations.

\begin{figure}[t]
  \centering
  \subfigure[MIMDP $\mdp$]{
    \scalebox{0.8}{
      \input{pics/mimdp_example}
    }
    \label{fig:mimdp_example}
    }
    \subfigure[Values for $\mdp$]{
    \scalebox{0.8}{
      \input{pics/mimdp_example_values}
    }
    \label{fig:mimdp_example_values}   
  }
  \caption{An example MIMDP and its possible valuations.}
\end{figure}
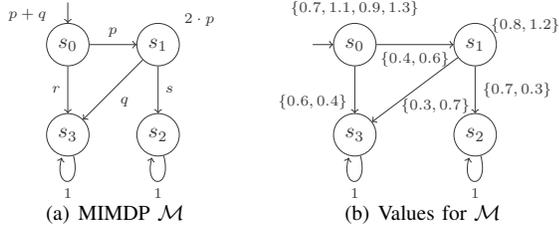
\begin{example}\label{ex:mimdp}
\figref{fig:mimdp_example} shows a MIMDP $\mdp$ with parametric transition probabilities $p,q,r$ and $s$. Costs are $p+q$ and $2\cdot p$.
Note that $\mdp$ contains no nondeterministic choices.
The valuations of the parameters are given by
\begin{align*}
	&\Valuations(p)=\{0.4,0.6\},\quad &\Valuations(q)=\{0.3,0.7\},\\
	&\Valuations(r)=\{0.6,0.4\},\quad &\Valuations(s)=\{0.7,0.3\},\\
	&\Valuations(p+q)=\{0.7,1.1,0.9,1.3\},\quad &\Valuations(2\cdot p)=\{0.8,1.2\}\ .
\end{align*}
These finite sets of valuations are also depicted in \figref{fig:mimdp_example_values} to indicate possible instantiations of the MIMDP. 
E.g., a valuation $u$ with $u(p)=0.4$ and $u(r)=0.4$ is \emph{not well--defined} as it induces no well--defined probability distribution.
\end{example}

\subsection*{Formal Problem Statement}
%
%
\begin{problem}\label{prob:fp}
For a MIMDP $\mdp$, a specification $\varphi=\reachProplT$, and
a set of goal states $G$
 , we want to find a valuation $u\in U(\Var)$ and a strategy $\sigma$ for the MDP $\mdp[u]$ such that $\mdp[u]^\sigma \models\varphi$, 
and the expected cost $\ec^{\mdp[u]^\sigma}(\finally G)$ is minimal. 
\end{problem}





%


%% file: pics/mimdp_example.tex
\begin{tikzpicture}[scale=1, nodestyle/.style={draw,circle},baseline=(s0)]
    
    \node [nodestyle,label={[label distance=0cm]130:\scriptsize$p+q$}] (s0) at (0,0) {$s_0$};
    \node [nodestyle,label={[label distance=0cm]30:\scriptsize$2\cdot p$}] (s1) [on grid, right= 1.5cm of s0] {$s_1$};
    \node [nodestyle] (s3) [on grid, below=1.5cm of s0] {$s_3$};
    \node [nodestyle] (s2) [on grid, below=1.5cm of s1] {$s_2$};
    
    \draw ($(s0)+(0,0.7)$) edge[->] (s0);
    \draw (s0) edge[->] node[auto] {\scriptsize$p$} (s1);
    \draw (s0) edge[->] node[left] {\scriptsize$r$} (s3);

    \draw (s1) edge[->] node[auto] {\scriptsize$s$} (s2);
    \draw (s1) edge[->] node[auto] {\scriptsize$q$} (s3);

    \draw (s3) edge[loop below, ->] node[auto] {\scriptsize$1$} (s3);
    \draw (s2) edge[loop below, ->] node[auto] {\scriptsize$1$} (s2);
    
\end{tikzpicture}

%% file: pics/mimdp_example_values.tex
\begin{tikzpicture}[scale=1, nodestyle/.style={draw,circle},baseline=(s0)]
    
    \node [nodestyle,label={[label distance=-0cm]90:\scriptsize$\{0.7,1.1,0.9,1.3\}$}] (s0) at (0,0) {$s_0$};
    \node [nodestyle,label={[label distance=-0.2cm]30:\scriptsize$\{0.8,1.2\}$}] (s1) [on grid, right=2cm of s0] {$s_1$};
    \node [nodestyle] (s3) [on grid, below=1.5cm of s0] {$s_3$};
    \node [nodestyle] (s2) [on grid, below=1.5cm of s1] {$s_2$};
    
    \draw ($(s0)-(0.7,0)$) edge[->] (s0);
    \draw (s0) edge[->] node[below] {\scriptsize$\{0.4,0.6\}$} (s1);
    \draw (s0) edge[->] node[left, near end] {\scriptsize$\{0.6,0.4\}$} (s3);

    \draw (s1) edge[->] node[auto] {\scriptsize$\{0.7,0.3\}$} (s2);
    \draw (s1) edge[->] node[right, near end] {\scriptsize$\{0.3,0.7\}$} (s3);

    \draw (s3) edge[loop below, ->] node[auto] {\scriptsize$1$} (s3);
    \draw (s2) edge[loop below, ->] node[auto] {\scriptsize$1$} (s2);    
\end{tikzpicture}

%% file: nonlinear.tex
\section{An Integer Programming Approach}
\noindent 
We first observe that Problem~\ref{prob:fp} is in fact a multi-objective verification problem an requires randomized strategies as in Def.~\ref{def:scheduler}~\cite{DBLP:conf/atva/ForejtKP12,DBLP:journals/lmcs/EtessamiKVY08,DBLP:conf/csl/BaierDK14}.
We first formulate a corresponding optimization problem using the following variables:
		\begin{itemize}
			\item $c_s\in\R_{\geq 0}$ for each $s\in S$
			represents the expected cost to reach $G\subseteq S$ from $s$.
			\item $p_s\in[0,1]$ for each $s\in S$ 
			represents the probability to reach $T\subseteq S$ from $s$.
			\item $\sched_s^\act\in[0,1]$ for each $s\in S$ and $\act\in\Act$
			represents the probability to choose action $\act\in\Act(s)$ at state $s$.
		\end{itemize}
We also introduce a characteristic variable $x_{u} \in \{0,1\}$ for each valuation  $u \in U(\Var)$.
If $x_u$ is set to $1$, all parameters and expressions are evaluated using $u$.
%
%
\begin{align}
			\text{minimize } &\quad c_{\sinit} \label{eq:min_structure}\\
			\text{subject to}\nonumber\\
										 &\quad p_{\sinit}\leq \lambda\label{eq:bound_structure_integer}\\
\forall s\in T.	 &\quad p_s=1\label{eq:target_structure_integer}\\
\forall s\in G.	 &\quad c_s=0\label{eq:goal_structure_integer}\\
&\quad \sum_{u \in U(\Var)} x_u =1 \label{eq:probOne_structure_integer} \\
\forall s\in S.&\quad p_s= \sum_{\act\in\Act(s)}\sched_s^\act\cdot \nonumber \\
& \quad\quad\quad \Bigl(\sum_{s'\in S} \sum_{u \in U(\Var)} x_{u} \cdot u(\probmdp(s,\act,s')) \cdot p_{s'}\Bigr)\label{eq:probcomputation_structure_integer}
\end{align}
\begin{align}
\forall s\in S.\quad c_s=&\sum_{\act\in\Act(s)}\sched_s^\act\cdot \Bigl(u(\rewFunction(s)) + \mbox{} \nonumber \\
& \sum_{s'\in S}\sum_{u \in U(\Var)} x_u \cdot u(\probmdp(s,\act,s')) \cdot u(\rewFunction(s'))\Bigr)\label{eq:rewcomputation_structure_integer}
\end{align}
\begin{align}
\forall s \in S.\,\forall\act\in\Act(s).\quad \sum_{s'\in S} \sum_{u \in U(\Var)} x_{u} \cdot u(\probmdp(s,\act,s')) =1 \label{eq:well-defined_probs_structure_integer}
\end{align}
\begin{theorem}[Soundness and completeness]\label{theo:correctness}
	The optimization problem~\eqref{eq:min_structure} -- \eqref{eq:well-defined_probs_structure_integer}  
	is \emph{sound} in the sense that each minimizing assignment induces a solution to Problem~\ref{prob:fp}. 	
	 It is \emph{complete} in the sense that for each solution to Problem~\ref{prob:fp} there is a minimizing assignment for~\eqref{eq:min_structure} -- \eqref{eq:well-defined_probs_structure_integer}.
\end{theorem}\smallskip

\paragraph{Proof Sketch.}
The first two equations induce satisfaction of the specifications:
\eqref{eq:min_structure} minimizes the expected cost to reach goal states $G\subseteq S$ at $\sinit$; \eqref{eq:bound_structure_integer} ensures that the probability to reach the target states $T\subseteq S$ from $\sinit$ is not higher than the threshold. 
\eqref{eq:target_structure_integer} and \eqref{eq:goal_structure_integer} set the probability and the expected cost at target and goal states to $1$ and $0$, respectively. 
\eqref{eq:probOne_structure_integer} ensures for all possibles values $u\in U(V)$ 
that exactly one corresponding characteristic variable $x_u$ is set to $1$.
In \eqref{eq:probcomputation_structure_integer}, $p_s$ is assigned the probability of reaching $T$ from $s$ by multiplying the probability to reach each successor $s'$ with the probability of reaching $T$ from $s'$, depending on the scheduler variables $\sched^\act_s$.
The variables $c_s$ are analogously assigned the expected cost to reach $G$ in \eqref{eq:rewcomputation_structure_integer}.
In \eqref{eq:well-defined_probs_structure_integer}, it is ensured that the concrete instantiations chosen at each transition form well--defined probability distributions.

Any satisfying assignment yields a well-defined randomized strategy and a well-defined assignment of parameters. 
Moreover, such an assignment necessarily satisfies the safety specification $\varphi=\reachProplT$, as the probability to reach $T$ is ensured to be smaller than or equal to $\lambda$. 
Likewise, the expected cost to reach $G$ from the initial state is minimized while at each state the $c_s$ variables are assigned the exact expected cost. 
Thus, a satisfying assignment induces a solution to Problem~\ref{prob:fp}.

\emph{Completeness} is given by construction, as all instantiations of Problem~\ref{prob:fp} may be encoded by~\eqref{eq:min_structure} -- \eqref{eq:well-defined_probs_structure_integer}.

\paragraph{Complexity of the Optimization Problem} Consider constraint~\eqref{eq:probcomputation_structure_integer}, where an integer variable $x_v$ is multiplied with the real--valued variable $p_{s'}$. Such constraints render this program a \emph{non--linear integer optimization problem} (NILP).
%
The number of constraints is governed by the number of state and action pairs or the number of possible instantiations of expressions \ie, the \emph{size of the problem} is in $\mathcal{O}(|S_r|\cdot |\Act| + |\Valuations(\mathcal{L}_\mdp)|^2)$.
The problem is, that already solving nonlinear problems without integer variables is NP-hard~\cite{bertsekas1999nonlinear,Las01}.
Summarized, despite the compact problem representation in form of a MIMDP, the problem is hard.

%% file: mdp_transformation.tex
\section{Transformation of \tool{PRISM} Programs}~\label{sec:program_trans}
\noindent We present a solution of Problem~\ref{prob:fp} that incorporates a direct transformation of MIMDPs specified as probabilistic programs in the \tool{PRISM} language as in Def.~\ref{def:prism_model}.
Similar to~\cite{quatmann-et-al-atva-2016}, we see the possible choices of parameter values as nondeterminism.
Say, a parameter $p\in \Var$ has valuations $\Valuations(p)=\{v_1,v_2\}$ and state $s$ has cost $\costFunction(s)=2\cdot p$. 
First, the MIMDP is transformed in the following way. 
From state $s$, a nondeterministic choice between actions $\act_{v_1}$ and $\act_{v_1}$ replaces the original transitions. 
Each action leads with probability one to a fresh state having cost $2\cdot v_1$ or $2\cdot v_2$, respectively. 
From these states, the original transitions of state $s$ emanate. 
Using a model checker to compute minimal or maximal expected cost in this transformed MDP allows to determine upper and lower bounds to a solution of Problem~\ref{prob:fp}.
The reason is, that we \emph{drop dependencies between parameters}. 
That is, if at one place $p$ is assigned its value $v_1$, it is not necessarily assigned the same everywhere in the MIMDP.

Then, we present a transformation that ensures parameter dependencies. 
Intuitively, in the resulting MDP each nondeterministic choice corresponding to a parameter value leads to a (sub-)MDP where the assignment of that value is fixed.

Note that if the original MIMDP has nondeterministic choices, we introduce a new level of nondeterminism. 
In that case, we assume that both types of nondeterminism minimize the expected cost for our problem. 
Alternatively, one can generate and evaluate a stochastic game~\cite{DBLP:conf/tacas/ChenFKPS13}.

\subsection*{Program Transformation 1---Parametric Cost} 
\noindent Assume a \tool{PRISM} program $M = (\Vars,\Act,C)$ as in Def.~\ref{def:prism_model}, and a parametric reward structure of the form:
\begin{lstlisting}[frame=tb]
rewards
 $g_1\colon l$
end rewards
\end{lstlisting}
with $g_1$ the guard and $l\in\expr(\Vars)$. Let $\Valuations(l)=\{\overline{v_1},\ldots,\overline{v_m}\}$ be the finite set of instantiations of $l$ with $\overline{v_i}\in\R$ for $1\leq i\leq m$. We introduce a fresh (characteristic) variable $x_l$ with $\dom(x_l)=\{0,\ldots,m\}$. Intuitively, there is a unique variable value for $x_l$ for each valuation from $\Valuations(l)$. 
Consider now all commands $c\in C$ of the form
 \begin{lstlisting}[frame=tb]
[$\act$] $g \rightarrow  p_1\colon f_1 + \ldots + p_n\colon f_n;$	
\end{lstlisting}
with $g\models g_1$, \ie, the guard of the command satisfies the guard of the reward structure. Replace each such commands $c$ by the following set of commands:
\begin{lstlisting}[frame=tb]
[] $g \rightarrow  1\colon x_l'=1;$
	$\vdots$
[] $g \rightarrow  1\colon x_l'=m;$
[$\act$] $\biglor\limits_{1\leq i\leq m}x_l=i\rightarrow p_1\colon f_1 + \ldots + p_n\colon f_n;$
\end{lstlisting}
and replace the reward structure for each command $c$ by
\begin{lstlisting}[frame=tb]
rewards
 $x_l=1\colon v_1;$
 	$\vdots$
 $x_l=m\colon v_m;$
end rewards
\end{lstlisting}
Intuitively, for each state satisfying the guard, $m$ transitions with the same guard are added.
Each of the transitions leads to new states with the instantiated rewards. From these states, the transitions of the original system emanate. This transformation corresponds to a nondeterministic choice between the concrete reward values.


\subsection*{Program Transformation 2---Parametric Transitions.}
\noindent For program $M = (\Vars,\Act,C)$, consider a command $c\in C$ of the form
 \begin{lstlisting}[frame=tb]
[$\act$] $g \rightarrow  p_1\colon f_1 + \ldots + p_n\colon f_n$	
\end{lstlisting}
with $p_1,\ldots,p_n\in\expr(\Vars)$. Let $\Valuations(p_1,\ldots,p_n)=\{v_1^n,\ldots,v_m^n\}$ with $v_i^n=(v_{i1},\ldots,v_{in})$ for $1\leq i\leq n$ and $v_{ij}\in\R$ for $1\leq j\leq m$.
 Replace each such command $c$ by the following set of commands:
\begin{lstlisting}[frame=tb]
[$\act$] $g \rightarrow v_{11}\colon f_1 + \ldots + v_{1n}\colon f_n ;$
	$\vdots$
[$\act$] $g \rightarrow v_{m1}\colon f_1 + \ldots + v_{mn}\colon f_n ;$
\end{lstlisting}
Intuitively, after the transformation, $m$ transitions with concrete probabilities emanate from the states.
As all transitions satisfy the same guard, again a nondeterministic choice between these transitions is induced.

For a program $M$, we denote the program after Transformation~1 and Transformation~2 by $M'$. The induced MIMDP of $M$ is denoted by $\mdp_M$ and the induced MDP of $M'$ by $\mdp_{M'}$.
\subsection*{Program Transformation 3---Parameter Dependencies}
\noindent We finally propose a transformation of the transformed MDP $\mdp_{M'}$ which enforces that once a parameter is assigned a specific value, this assignment is always used. 
Therefore, we add a \emph{control module} to the PRISM formulation.

In the transformed MDP, taking actions $\act_{v_1}$ or $\act_{v_2}$ induces that parameter $p$ is assigned its value $v_1$ or $v_2$. In the PRISM encoding, the corresponding commands are of the form

\begin{lstlisting}[frame=tb]
[$\act_{v_1}$] $g_1 \rightarrow  \ldots;$
[$\act_{v_2}$] $g_2 \rightarrow  \ldots;$
\end{lstlisting}
where $g_1$ and $g_2$ are arbitrary guards.
Now, for each of these actions $\act_{v_1}$ and $\act_{v_2}$, we use control variables $q_{v_1}$ and $q_{v_2}$ and build a control module of the form:

\begin{lstlisting}[frame=tb]
module control
  $q_{v_1}$: bool init $0$;
  $q_{v_2}$: bool init $0$;
  [$\act_{v_1}$] $\neg q_{v_1} \rightarrow (q_{v_1}'=\true);$
  [$\act_{v_1}$] $q_{v_1} \rightarrow (q_{v_1}'=\true);$
  [$\act_{v_2}$] $\neg q_{v_2} \rightarrow (q_{v_2}'=\true);$
  [$\act_{v_2}$] $q_{v_2} \rightarrow (q_{v_2}'=\true);$
endmodule
\end{lstlisting}
If this module is included in the parallel composition, a control variable is set to \true once the corresponding action is taken in the MDP. 
%
We can now guard the commands such that only non--conflicting assignments are possible. The original commands are transformed in the following way:
\begin{lstlisting}[frame=tb]
[$\act_{v_1}$] $g_1\land\neg q_{v_2} \rightarrow  \ldots;$
[$\act_{v_2}$] $g_2\land\neg q_{v_1} \rightarrow  \ldots;$
\end{lstlisting}
In the resulting \emph{controlled MDP}, it is not possible that the original parameter $p$ is at one place assigned $v_1$ and at another place $v_2$. 

Basically, we transform an integer nonlinear optimization problem to a linear program at the cost of increasing the size of the underlying MDP. 
%
We denote the program $M$ subject to Transformation~3 by $M_c$.
Given a program defining a MIMDP $\mdp$ which is subject to Transformations~1 -- 3, we denote the resulting program by $M'_c$ and the induced MDP by $\mdp_{M'_c}$.

\begin{proposition}
  Given the MDP $\mdp_{M'_c}$, a \emph{reachability property} $\varphi=\reachProplT$, and a set of goal states $G$.
  Finding a strategy $\sched\in\Sched^{\mdp_{M'_c}}$ that induces the minimal expected cost $\ec^{\mdp_{M'_c}^{\sched}}$ under all strategies that satisfy $\mdp_{M'_c}^{\sched}\models\reachPropSymbol$ provides a solution to Problem~1.	
\end{proposition}
The correctness of that proposition is given by the step-by-step transformation provided above. 
Basically, introducing a nondeterministic choices for all parameter values is nothing else then trying out all these values. 
With the control module, consistent choices are enforced.
Intuitively, the first nondeterministic choice corresponding to a specific parameter value leads to (sub-)MDP that allows only that value.
We exploit highly optimized and specialized model checking tools.
Aggressive state space reduction techniques together with a preprocessing that removes inconsistent  combinations of parameter values beforehand render the MIMDP synthesis problem feasible.



%

%% file: experiments.tex
\section{Experiments}\label{sec:experiments}
We first report on results for the case study from Sec.~\ref{sec:casestudy}.
First, we created a \tool{PRISM} program for the MIMDP underlying all aspects of the shipyard example with 430 lines of code.
We can use the program to generate an explicit Markov chain (MC) model where the parameter instantiations are fixed. 
That explicit model has 1\,728 states and 5\,145 transitions.
From the MIMDP model, we generate the MDP according to the transformations in Sec.~\ref{sec:program_trans}. 
The underlying (extended) \tool{PRISM} program has 720 lines of code. 
The explicit MDP generated from the transformed program has 2\,912 states and 64\,048 transitions.
For our case study, the size of the transformed MDP is reasonable:
From the MIMDP to the MDP, states increase by a factor of 1.8, transitions by a factor of 12.5.

\begin{figure}
	\centering
	\scalebox{0.75}{\input{plot_recogError_moreMissions_quadratic}}
	\caption{Recognition Error Probabilities}
	\label{fig:plot_recognition_error}
\end{figure}
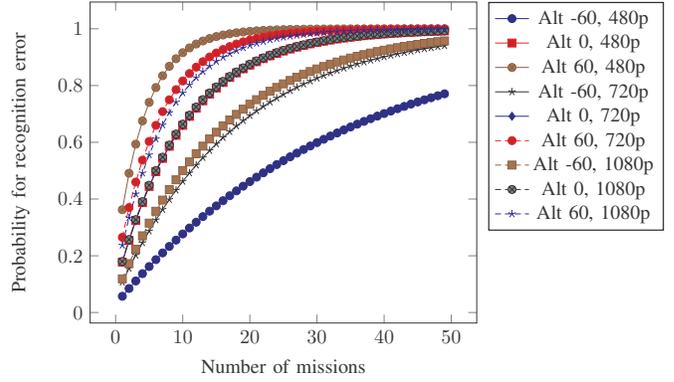

The experimental results show several (partially unforeseeable) intricacies of the case study.
We have the following structure for the MIMDP defined by parameters and their valuation sets. 
For details see Sec.~\ref{sec:casestudy}.
\begin{itemize}
	\item EO sensor for the UAV: $V_{\text{EO}}=\{480p,720p,1080p\}$.
	\item Deviation from the UAV operational altitude: $V_{\text{Alt}}\{-60,-30,0,30,60\}$.
	\item ROC ground sensors (Sensor 1, Sensor 2, Sensor 3, Bay Area Network):  $V_{\text{ROC}}=\{\text{low},\text{med},\text{high}\}$.
	\item False positive rates for each ROC ground sensor: $V_{\text{fp}}=\{0.2,0.1,\ldots,1.0\}$.
\end{itemize}
This structure induces $1\,440$ possibilities to configure the system.
However, as we restricted all ground sensors to have the same quality, we have only $360$ possibilities.
We implemented several benchmark scripts using the \tool{Python} interface of the \tool{storm} model checker. 
For each of the results presented below, we used a script to iteratively try all possible combinations for measures of interest, and compared the optimal value obtained from the transformed MDP to these results.
We performed the experiments on a MacBook Pro with a 2.3 GHz Intel Core i5 CPU and 8GB of RAM.
\paragraph*{Probability of Recognition Error}
%
%
%
%
First, we investigate the probability of not recognizing an intruder---after a ground sensor was triggered---in dependence of the number of missions an UAV flies.
The curves shown in Fig.~\ref{fig:plot_recognition_error} depend on the deviation from the standard UAV operational altitude and on the type of EO sensor.
The cheap 480p EO sensor has the lowest probability for an recognition error at a low altitude.
For all other sensors and altitudes, the probability quickly approaches one.
The cumulated model checking time was $78.13$ seconds, computing the optimal result on the transformed MDP took $2.3$ seconds.
	
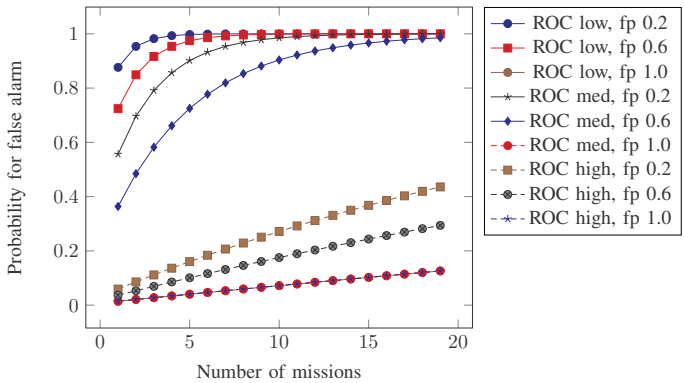
\begin{figure}
	\centering
	\scalebox{0.75}{\input{plot_falseAlarm_all_missioncount}}
	\caption{False Alarm Probabilities}
	\label{fig:plot_false_alarm}
\end{figure}

\paragraph*{Probability of False Alarms}
For the ROC sensors, we measured the probability for a false alarm, depending again on the number of missions.
The curves in Fig.~\ref{fig:plot_false_alarm} depend on the quality of the ROC sensor and on the false positive rates. 
The high-quality sensor is the only one with relatively low false alarm probabilities.
The cumulated model checking time was $52.13$ seconds, computing the optimal result on the transformed MDP took less than one second.

%
%
%

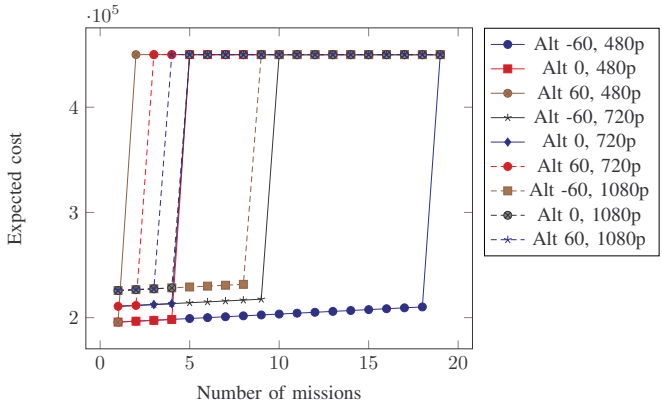
\begin{figure}
	\scalebox{0.75}{\input{plot_expected_cost_vs_prob_recog.tex}}
	\caption{Probability and Cost Tradeoffs}
	\label{fig:plot_tradeoffs}
\end{figure}

\paragraph*{Probability and Cost Tradeoffs}
Finally, we show the expected cost in dependence of the number of missions in Fig.~\ref{fig:plot_tradeoffs}. 
Additionally, for each data point, the probability for a recognition error needs to be below $50\%$ .
The violation of this property is indicated by the maximum value $4.5\cdot 10^{-5}$.
The results show that for this kind of property indeed the low-resolution (480p) sensor at lowest altitude is the best choice, as it has relatively low initial cost.
While the task cost at low altitude is slightly larger than at higher altitudes, with this sensor, the UAV is able to maintain the probability threshold of safely recognizing an intruder.
The cumulated model checking time was $59.1$ seconds, computing the optimal result on the transformed MDP took $1.2$ seconds.
	
\subsection*{\prism Benchmarks}
\begin{table}[]
\centering
\def\arraystretch{0.7}
\setlength{\tabcolsep}{0.55em} 
\caption{Parametric Benchmarks}
\label{tab:benchmarks}
\begin{tabular}{@{}|l|l|r|r|r|l|@{}}
\toprule
Model                    & Type        & States & Transitions & MC (s) & SE (s) \\ \midrule
\multirow{3}{*}{Die}     & parametric  & 13     & 20    & ---    & 0.04       \\ \cmidrule(l){2-6} 
                         & transformed & 13     & 48    & 0.04      & ---    \\ \cmidrule(l){2-6} 
                         & controlled  & 37     & 60    & 0.02      & ---    \\ \midrule
\multirow{3}{*}{Zeroconf1} & parametric  & 1\,004   & 2\,005  & ---    & 60.3       \\ \cmidrule(l){2-6} 
                         & transformed & 1\,004   & 6\,009  & 0.18      & ---    \\ \cmidrule(l){2-6} 
                         & controlled  & 9\,046   & 18\,075 & 0.19      & ---    \\ \midrule
\multirow{3}{*}{Zeroconf2} & parametric  & 10\,004   & 20\,005  & ---    & TO       \\ \cmidrule(l){2-6} 
                         & transformed & 10\,004   & 60\,009  & 0.11     & ---    \\ \cmidrule(l){2-6} 
                         & controlled  & 90\,046   & 180\,075 & 0.59      & ---    \\ \midrule
\multirow{3}{*}{Zeroconf3} & parametric  & 100\,004   & 200\,005  & ---    & TO      \\ \cmidrule(l){2-6} 
                         & transformed & 100\,004   & 600\,009  &  0.90      & ---    \\ \cmidrule(l){2-6} 
                         & controlled  & 900\,046   & 1\,800\,075 & 7.36      & ---    \\ \midrule
\multirow{3}{*}{Crowds1} & parametric  & 1\,367   & 2\,027  & ---    & 0.97      \\ \cmidrule(l){2-6} 
                         & transformed & 1\,367   & 3\,347  &  0.04      & ---    \\ \cmidrule(l){2-6} 
                         & controlled  & 12\,328   & 18\,378 & 0.11      & ---    \\ \midrule
\multirow{3}{*}{Crowds2} & parametric  & 7\,421   & 12\,881  & ---    & 5.80      \\ \cmidrule(l){2-6} 
                         & transformed & 7\,421   & 20\,161  &  0.08      & ---    \\ \cmidrule(l){2-6} 
                         & controlled  & 66\,826   & 116\,148 & 0.42      & ---    \\ \midrule
\multirow{3}{*}{Crowds3} & parametric  & 104\,512   & 246\,082  & ---    &  128.62     \\ \cmidrule(l){2-6} 
                         & transformed & 104\,512   & 349\,042  &  0.74      & ---    \\ \cmidrule(l){2-6} 
                         & controlled  & 940\,675   & 2\,215\,167 & 6.29      & ---    \\ \midrule
\multirow{4}{*}{Crowds4} & parametric  & 572\,153   & 1\,698\,233  & ---    &  18.81     \\ \cmidrule(l){2-6} 
                         & transformed & 572\,153   & 2\,261\,273  &  4.39      & ---    \\ \cmidrule(l){2-6} 
                         & controlled  & 5\,149\,474   & 15\,284\,736 & 33.70      & ---    \\ \midrule
\multirow{4}{*}{Crowds5} & parametric  & 2\,018\,094   & 7\,224\,834  & ---    &  TO  \\ \cmidrule(l){2-6} 
                         & transformed & 2\,018\,094   & 9\,208\,354  &  17.15      & ---    \\ \cmidrule(l){2-6} 
                         & controlled  & 18\,162\,973   & 65\,024\,355 & 137.71  & ---    \\ \bottomrule                                                                                                
\end{tabular}
\end{table}

We additionally assessed the feasibility of our approach on well-known case studies, parametric Markov chain examples from the \tool{PARAM}-website~\cite{param_website}, that originally stem from the \tool{PRISM} benchmark suite~\cite{KNP12b}. 
We tested a parametric version of the Knuth-Yao Die (Die), several instances of the Zeroconf protocol~\cite{Zeroconf03}, and the Crowds protocol~\cite{RR98}.
For benchmarking, we used the standard configuration of \tool{storm}.
Table~\ref{tab:benchmarks} shows the results, where we list the number of states and transitions as well as the model checking times (MC):
``transformed'' refers to the MDP after Transformation~1 and~2, ``controlled refers to the MDP after Transformations~1 -- 3.
For the original parametric Markov chains, we tested the time to perform state elimination, which is the standard model checking method for such models~\cite{param_sttt,daws04}.
We used a timeout (TO) of 600 seconds.
From the results, we draw the following conclusions: (1) The first two program transformations only increase the number of transitions with respect to the MIMDP model, and the third transformation increases the states and transitions by up to one order of magnitude. (2) Except for the Crowds4 instance, model checking of the transformed and controlled MDP is by far superior to performing state elimination.
(3) For these standard benchmarks, we are able to handle instances with millions of states.

%% file: plot_recogError_moreMissions_quadratic.tex
\begin{tikzpicture}
 \begin{axis}[
 xlabel={Number of missions},
 ylabel={Probability for recognition error},
 legend pos=outer north east]\addplot coordinates {
(1,0.0572517)(2,0.0846338)(3,0.11122)(4,0.137035)(5,0.162099)(6,0.186435)(7,0.210065)(8,0.233008)(9,0.255284)(10,0.276914)(11,0.297915)(12,0.318307)(13,0.338106)(14,0.35733)(15,0.375996)(16,0.39412)(17,0.411717)(18,0.428803)(19,0.445393)(20,0.461501)(21,0.477142)(22,0.492328)(23,0.507073)(24,0.521389)(25,0.53529)(26,0.548787)(27,0.561892)(28,0.574617)(29,0.586972)(30,0.598968)(31,0.610615)(32,0.621925)(33,0.632906)(34,0.643568)(35,0.65392)(36,0.663971)(37,0.673731)(38,0.683207)(39,0.692408)(40,0.701342)(41,0.710016)(42,0.718439)(43,0.726616)(44,0.734557)(45,0.742266)(46,0.749752)(47,0.75702)(48,0.764077)(49,0.770929)
};
\addlegendentry{Alt -60, 480p}

\addplot coordinates {
(1,0.178006)(2,0.254741)(3,0.324313)(4,0.387389)(5,0.444577)(6,0.496426)(7,0.543435)(8,0.586055)(9,0.624697)(10,0.659732)(11,0.691496)(12,0.720295)(13,0.746405)(14,0.770078)(15,0.791542)(16,0.811001)(17,0.828644)(18,0.84464)(19,0.859143)(20,0.872292)(21,0.884214)(22,0.895022)(23,0.904822)(24,0.913707)(25,0.921762)(26,0.929066)(27,0.935688)(28,0.941691)(29,0.947134)(30,0.952069)(31,0.956544)(32,0.9606)(33,0.964278)(34,0.967613)(35,0.970636)(36,0.973377)(37,0.975863)(38,0.978116)(39,0.980159)(40,0.982011)(41,0.98369)(42,0.985213)(43,0.986593)(44,0.987845)(45,0.988979)(46,0.990008)(47,0.990941)(48,0.991787)(49,0.992553)
};
\addlegendentry{Alt 0, 480p}

\addplot coordinates {
(1,0.362333)(2,0.490787)(3,0.593364)(4,0.675277)(5,0.740689)(6,0.792924)(7,0.834637)(8,0.867947)(9,0.894548)(10,0.91579)(11,0.932753)(12,0.946299)(13,0.957116)(14,0.965755)(15,0.972653)(16,0.978162)(17,0.982561)(18,0.986074)(19,0.988879)(20,0.991119)(21,0.992908)(22,0.994337)(23,0.995478)(24,0.996389)(25,0.997116)(26,0.997697)(27,0.998161)(28,0.998531)(29,0.998827)(30,0.999063)(31,0.999252)(32,0.999403)(33,0.999523)(34,0.999619)(35,0.999696)(36,0.999757)(37,0.999806)(38,0.999845)(39,0.999876)(40,0.999901)(41,0.999921)(42,0.999937)(43,0.99995)(44,0.99996)(45,0.999968)(46,0.999974)(47,0.99998)(48,0.999984)(49,0.999987)
};
\addlegendentry{Alt 60, 480p}

\addplot coordinates {
(1,0.106703)(2,0.155701)(3,0.202011)(4,0.245781)(5,0.287149)(6,0.326249)(7,0.363204)(8,0.398132)(9,0.431144)(10,0.462345)(11,0.491835)(12,0.519708)(13,0.546051)(14,0.57095)(15,0.594483)(16,0.616725)(17,0.637748)(18,0.657617)(19,0.676397)(20,0.694146)(21,0.710922)(22,0.726778)(23,0.741764)(24,0.755928)(25,0.769315)(26,0.781968)(27,0.793927)(28,0.80523)(29,0.815913)(30,0.82601)(31,0.835553)(32,0.844573)(33,0.853098)(34,0.861155)(35,0.868771)(36,0.875969)(37,0.882772)(38,0.889202)(39,0.895279)(40,0.901023)(41,0.906452)(42,0.911583)(43,0.916432)(44,0.921016)(45,0.925348)(46,0.929443)(47,0.933313)(48,0.93697)(49,0.940428)
};
\addlegendentry{Alt -60, 720p}

\addplot coordinates {
(1,0.177868)(2,0.254553)(3,0.324085)(4,0.387131)(5,0.444297)(6,0.496129)(7,0.543128)(8,0.585742)(9,0.624381)(10,0.659417)(11,0.691184)(12,0.719989)(13,0.746107)(14,0.769788)(15,0.791261)(16,0.810731)(17,0.828385)(18,0.844392)(19,0.858906)(20,0.872066)(21,0.883999)(22,0.894819)(23,0.90463)(24,0.913525)(25,0.921591)(26,0.928905)(27,0.935536)(28,0.941549)(29,0.947001)(30,0.951944)(31,0.956427)(32,0.960491)(33,0.964176)(34,0.967517)(35,0.970547)(36,0.973294)(37,0.975785)(38,0.978044)(39,0.980092)(40,0.981949)(41,0.983632)(42,0.985159)(43,0.986543)(44,0.987799)(45,0.988937)(46,0.989969)(47,0.990904)(48,0.991753)(49,0.992522)
};
\addlegendentry{Alt 0, 720p}

\addplot coordinates {
(1,0.265096)(2,0.369983)(3,0.4599)(4,0.536983)(5,0.603064)(6,0.659715)(7,0.70828)(8,0.749914)(9,0.785605)(10,0.816203)(11,0.842435)(12,0.864922)(13,0.8842)(14,0.900727)(15,0.914895)(16,0.927041)(17,0.937454)(18,0.94638)(19,0.954033)(20,0.960593)(21,0.966217)(22,0.971039)(23,0.975172)(24,0.978715)(25,0.981753)(26,0.984357)(27,0.98659)(28,0.988504)(29,0.990144)(30,0.991551)(31,0.992757)(32,0.993791)(33,0.994677)(34,0.995436)(35,0.996088)(36,0.996646)(37,0.997125)(38,0.997535)(39,0.997887)(40,0.998188)(41,0.998447)(42,0.998669)(43,0.998859)(44,0.999022)(45,0.999161)(46,0.999281)(47,0.999384)(48,0.999472)(49,0.999547)
};
\addlegendentry{Alt 60, 720p}

\addplot coordinates {
(1,0.118518)(2,0.172395)(3,0.222979)(4,0.270471)(5,0.31506)(6,0.356924)(7,0.396228)(8,0.433131)(9,0.467778)(10,0.500307)(11,0.530848)(12,0.559523)(13,0.586445)(14,0.611721)(15,0.635453)(16,0.657734)(17,0.678653)(18,0.698293)(19,0.716734)(20,0.734047)(21,0.750302)(22,0.765563)(23,0.779892)(24,0.793345)(25,0.805976)(26,0.817834)(27,0.828968)(28,0.839422)(29,0.849236)(30,0.858451)(31,0.867102)(32,0.875225)(33,0.882851)(34,0.890011)(35,0.896734)(36,0.903045)(37,0.908971)(38,0.914535)(39,0.919759)(40,0.924663)(41,0.929267)(42,0.933591)(43,0.93765)(44,0.94146)(45,0.945038)(46,0.948398)(47,0.951551)(48,0.954513)(49,0.957293)
};
\addlegendentry{Alt -60, 1080p}

\addplot coordinates {
(1,0.1799)(2,0.257316)(3,0.327423)(4,0.390912)(5,0.448408)(6,0.500476)(7,0.547629)(8,0.590331)(9,0.629002)(10,0.664022)(11,0.695737)(12,0.724458)(13,0.750468)(14,0.774023)(15,0.795354)(16,0.814672)(17,0.832166)(18,0.848008)(19,0.862356)(20,0.875349)(21,0.887115)(22,0.897771)(23,0.907421)(24,0.91616)(25,0.924074)(26,0.931241)(27,0.937732)(28,0.94361)(29,0.948933)(30,0.953753)(31,0.958119)(32,0.962072)(33,0.965652)(34,0.968894)(35,0.971831)(36,0.97449)(37,0.976898)(38,0.979079)(39,0.981053)(40,0.982842)(41,0.984462)(42,0.985928)(43,0.987257)(44,0.98846)(45,0.989549)(46,0.990535)(47,0.991429)(48,0.992238)(49,0.992971)
};
\addlegendentry{Alt 0, 1080p}

\addplot coordinates {
(1,0.237105)(2,0.33365)(3,0.417977)(4,0.491631)(5,0.555964)(6,0.612156)(7,0.661237)(8,0.704106)(9,0.741551)(10,0.774257)(11,0.802824)(12,0.827776)(13,0.84957)(14,0.868607)(15,0.885234)(16,0.899758)(17,0.912443)(18,0.923523)(19,0.933201)(20,0.941654)(21,0.949038)(22,0.955487)(23,0.96112)(24,0.96604)(25,0.970338)(26,0.974091)(27,0.97737)(28,0.980234)(29,0.982735)(30,0.98492)(31,0.986828)(32,0.988495)(33,0.989951)(34,0.991223)(35,0.992333)(36,0.993304)(37,0.994151)(38,0.994891)(39,0.995538)(40,0.996102)(41,0.996596)(42,0.997026)(43,0.997403)(44,0.997731)(45,0.998018)(46,0.998269)(47,0.998488)(48,0.99868)(49,0.998847)
};
\addlegendentry{Alt 60, 1080p}

\end{axis}
 \end{tikzpicture}

%% file: plot_falseAlarm_all_missioncount.tex
\begin{tikzpicture}
 \begin{axis}[
 xlabel={Number of missions},
 ylabel={Probability for false alarm},
 legend pos=outer north east]\addplot coordinates {
(1,0.876318)(2,0.953669)(3,0.982217)(4,0.993204)(5,0.9974)(6,0.999005)(7,0.999619)(8,0.999854)(9,0.999944)(10,0.999979)(11,0.999992)(12,0.999997)(13,0.999999)(14,1)(15,1)(16,1)(17,1)(18,1)(19,1)
};
\addlegendentry{ROC low, fp 0.2}

\addplot coordinates {
(1,0.724156)(2,0.848913)(3,0.916553)(4,0.953961)(5,0.97459)(6,0.985975)(7,0.992258)(8,0.995726)(9,0.997641)(10,0.998698)(11,0.999281)(12,0.999603)(13,0.999781)(14,0.999879)(15,0.999933)(16,0.999963)(17,0.99998)(18,0.999989)(19,0.999994)
};
\addlegendentry{ROC low, fp 0.6}

\addplot coordinates {
(1,0.0142587)(2,0.0208539)(3,0.0273751)(4,0.0338543)(5,0.0402893)(6,0.0466811)(7,0.0530301)(8,0.0593368)(9,0.0656015)(10,0.0718244)(11,0.0780058)(12,0.0841461)(13,0.0902455)(14,0.0963042)(15,0.102323)(16,0.108301)(17,0.11424)(18,0.120138)(19,0.125998)
};
\addlegendentry{ROC low, fp 1.0}

\addplot coordinates {
(1,0.556312)(2,0.696386)(3,0.7915)(4,0.856867)(5,0.901726)(6,0.932523)(7,0.953668)(8,0.968186)(9,0.978154)(10,0.984999)(11,0.9897)(12,0.992927)(13,0.995143)(14,0.996665)(15,0.99771)(16,0.998427)(17,0.99892)(18,0.999259)(19,0.999491)
};
\addlegendentry{ROC med, fp 0.2}

\addplot coordinates {
(1,0.363584)(2,0.484639)(3,0.582068)(4,0.661114)(5,0.725194)(6,0.777153)(7,0.819286)(8,0.853452)(9,0.881159)(10,0.903627)(11,0.921847)(12,0.936622)(13,0.948604)(14,0.958321)(15,0.966201)(16,0.972591)(17,0.977772)(18,0.981975)(19,0.985383)
};
\addlegendentry{ROC med, fp 0.6}

\addplot coordinates {
(1,0.0142587)(2,0.0208539)(3,0.0273751)(4,0.0338543)(5,0.0402893)(6,0.0466811)(7,0.0530301)(8,0.0593368)(9,0.0656015)(10,0.0718244)(11,0.0780058)(12,0.0841461)(13,0.0902455)(14,0.0963042)(15,0.102323)(16,0.108301)(17,0.11424)(18,0.120138)(19,0.125998)
};
\addlegendentry{ROC med, fp 1.0}

\addplot coordinates {
(1,0.0592205)(2,0.085682)(3,0.111278)(4,0.136164)(5,0.160349)(6,0.183855)(7,0.206703)(8,0.228911)(9,0.250496)(10,0.271478)(11,0.291872)(12,0.311695)(13,0.330964)(14,0.349693)(15,0.367897)(16,0.385592)(17,0.402792)(18,0.41951)(19,0.43576)
};
\addlegendentry{ROC high, fp 0.2}

\addplot coordinates {
(1,0.0364496)(2,0.0530274)(3,0.0692447)(4,0.0851879)(5,0.100855)(6,0.116254)(7,0.131388)(8,0.146263)(9,0.160883)(10,0.175252)(11,0.189376)(12,0.203257)(13,0.216901)(14,0.230312)(15,0.243492)(16,0.256447)(17,0.26918)(18,0.281695)(19,0.293996)
};
\addlegendentry{ROC high, fp 0.6}

\addplot coordinates {
(1,0.0142587)(2,0.0208539)(3,0.0273751)(4,0.0338543)(5,0.0402893)(6,0.0466811)(7,0.0530301)(8,0.0593368)(9,0.0656015)(10,0.0718244)(11,0.0780058)(12,0.0841461)(13,0.0902455)(14,0.0963042)(15,0.102323)(16,0.108301)(17,0.11424)(18,0.120138)(19,0.125998)
};
\addlegendentry{ROC high, fp 1.0}

\end{axis}
 \end{tikzpicture}

%% file: plot_expected_cost_vs_prob_recog.tex
\begin{tikzpicture}
 \begin{axis}[
 xlabel={Number of missions},
 ylabel={Expected cost},
 legend pos=outer north east]\addplot coordinates {
(1,195819)(2,196650)(3,197487)(4,198328)(5,199172)(6,200017)(7,200863)(8,201710)(9,202557)(10,203404)(11,204252)(12,205099)(13,205947)(14,206794)(15,207641)(16,208489)(17,209336)(18,210184)(19,450000)
};
\addlegendentry{Alt -60, 480p}

\addplot coordinates {
(1,195806)(2,196625)(3,197450)(4,198279)(5,450000)(6,450000)(7,450000)(8,450000)(9,450000)(10,450000)(11,450000)(12,450000)(13,450000)(14,450000)(15,450000)(16,450000)(17,450000)(18,450000)(19,450000)
};
\addlegendentry{Alt 0, 480p}

\addplot coordinates {
(1,195793)(2,450000)(3,450000)(4,450000)(5,450000)(6,450000)(7,450000)(8,450000)(9,450000)(10,450000)(11,450000)(12,450000)(13,450000)(14,450000)(15,450000)(16,450000)(17,450000)(18,450000)(19,450000)
};
\addlegendentry{Alt 60, 480p}

\addplot coordinates {
(1,210816)(2,211644)(3,212478)(4,213316)(5,214157)(6,214999)(7,215842)(8,216686)(9,217530)(10,450000)(11,450000)(12,450000)(13,450000)(14,450000)(15,450000)(16,450000)(17,450000)(18,450000)(19,450000)
};
\addlegendentry{Alt -60, 720p}

\addplot coordinates {
(1,210802)(2,211619)(3,212441)(4,213267)(5,450000)(6,450000)(7,450000)(8,450000)(9,450000)(10,450000)(11,450000)(12,450000)(13,450000)(14,450000)(15,450000)(16,450000)(17,450000)(18,450000)(19,450000)
};
\addlegendentry{Alt 0, 720p}

\addplot coordinates {
(1,210790)(2,211594)(3,450000)(4,450000)(5,450000)(6,450000)(7,450000)(8,450000)(9,450000)(10,450000)(11,450000)(12,450000)(13,450000)(14,450000)(15,450000)(16,450000)(17,450000)(18,450000)(19,450000)
};
\addlegendentry{Alt 60, 720p}

\addplot coordinates {
(1,225812)(2,226636)(3,227466)(4,228301)(5,229137)(6,229976)(7,230815)(8,231654)(9,450000)(10,450000)(11,450000)(12,450000)(13,450000)(14,450000)(15,450000)(16,450000)(17,450000)(18,450000)(19,450000)
};
\addlegendentry{Alt -60, 1080p}

\addplot coordinates {
(1,225799)(2,226613)(3,227432)(4,228255)(5,450000)(6,450000)(7,450000)(8,450000)(9,450000)(10,450000)(11,450000)(12,450000)(13,450000)(14,450000)(15,450000)(16,450000)(17,450000)(18,450000)(19,450000)
};
\addlegendentry{Alt 0, 1080p}

\addplot coordinates {
(1,225787)(2,226587)(3,227394)(4,450000)(5,450000)(6,450000)(7,450000)(8,450000)(9,450000)(10,450000)(11,450000)(12,450000)(13,450000)(14,450000)(15,450000)(16,450000)(17,450000)(18,450000)(19,450000)
};
\addlegendentry{Alt 60, 1080p}

\end{axis}
 \end{tikzpicture}

%% file: outlook.tex
%

\section{Conclusion and Future Work}
We introduced the structured synthesis problem for Markov decision processes.
Driven by a concrete case study from the area of physical security, we defined multiple-instance MDPs and demonstrated the hardness of the corresponding synthesis problem. 
As a feasible solution, we presented a transformation to an MDP where nondeterministic choices represented the underlying structure of the multiple-instance MDP. 
Our experiments showed that we are able to analyze meaningful measures for such problems.
In the future, we will further investigate intricacies of the case study, for instance regarding continuous state spaces.
Moreover, we will extend our methodology to models based on reinforcement learning~\cite{sut98a}, also considering safety aspects~\cite{junges-et-al-tacas-2016,garcia2015comprehensive}.

%% file: appendix.tex
\newpage\appendix

\section{Case Study--Details}\label{app:casestudy}
\input{case_study_details}

%% file: case_study_details.tex
\subsection{UAV Task Distances}
\label{s:distances}
\begin{table}[ht]
\caption{Task distances in meters for point search tasks.}
\centering
\begin{tabular}{| l | c | c | }
\hline
 \textbf{Task} & $d_g$ & $d_g'$ \\
 \hline
Main Gate & 1000 & 1000 \\
 \hline
Power Generator & 1100 & 1100 \\
 \hline
Sensor 1 & 3250 & 3250 \\
 \hline
Sensor 2 & 3375 & 3375 \\
 \hline
Sensor 3 & 2375 & 2375 \\
 \hline
Bay Sensor Network & 3750 & 3750 \\
\hline
\end{tabular}
\label{t:pointTaskDims}
\end{table}
For a point search task $t_p \in \{$Main Gate, Power Generator, Sensor 1, Sensor~2, Sensor 3, Bay Sensor Network\}, 
let $d_g(t_p)$ be the distance from the Ground Control Station to the task location  
and $d_g'(t_p)$ be the distance back, as given in \tabref{t:pointTaskDims}.
Then the total distance $d(t_p)$ traveled to perform task $t_p$ is 
\begin{equation}
d(t_p) = d_g(t_p) + d_g'(t_p).
\label{eq:pointDistances}
\end{equation}
In this case distances are symmetric, with $d_g(t_p) = d_g'(t_p)$ for all $t_p$.

\begin{table}[ht]
\caption{Task distances in meters for line search tasks.}
\centering
\begin{tabular}{| l | c | c | c | }
\hline
 \textbf{Task} & $d_g$ & $d_g'$ & $d_l$ \\
 \hline
Highway & 2000 & 2000 & 2875 \\
 \hline
Runway & 1000 & 1250 & 2000 \\
 \hline
Stream & 3250 & 3250 & 2050 \\
\hline
\end{tabular}
\label{t:lineTaskDims}
\end{table}
For a line search task $t_l \in \{$Highway, Runway, Stream\}, 
let $d_g(t_l)$ be the distance from the Ground Control Station to the start of the line, 
$d_g'(t_l)$ the distance back to the Ground Control Station from the end of the line, 
and $d_l(t_l)$ the length of the line to be searched, 
as given in \tabref{t:lineTaskDims}.
Then the total distance $d(t_l)$ traveled to perform task $t_l$ is 
\begin{equation}
d(t_l) = d_g(t_l) + d_l(t_l)+ d_g'(t_l).
\label{eq:lineDistances}
\end{equation}

\begin{table*}[ht]
\caption{Base task distances in meters for area search tasks.}
\centering
\begin{tabular}{| l | c | c | c | c | c | c  |}
\hline
 \textbf{Task $t_a$} & $d_g(t_a)$ & $d_g'(t_a)$ & $d_s(t_a)$ & $d_s'(t_a)$ & $d_h(t_a)$  & $d_w(t_a)$ \\
 \hline
Bridge 		& 3750 & 3500 		& $d_g(\textrm{Sensor 1}) = 3250$  & 500 		& 375 & 600 \\
 \hline
Main Office 	& 2600 & 1525 		& $d_g(\textrm{Sensor 1}) = 3250$  & 625 		& 500 & 1100 \\
 \hline
Warehouse 	& 2750 & 1250 		& $d_g(\textrm{Sensor 1}) = 3250$  & 575  		& 700 & 1500 \\
 
 \hline
Truck Depot 	& 2950 & 2100 		& $d_g(\textrm{Sensor 2}) = 3375$ & 	450			& 625 & 1200 \\
 \hline
Shipyard Office & 3350 & 3150 		& $d_g(\textrm{Sensor 2}) = 3375$ &  425			& 550 & 1000 \\
 \hline 
Shipyard 		& 3625 & 3050 		& \{$d_g(\textrm{Sensor 2}), d_g(\textrm{Bay Sensor})$\} & 500		& 800 & 700 \\
 
 \hline
West Bay 		& 4250 & 4900 		& $d_g(\textrm{Bay Sensor}) = 3750$ & 	1000			& 575 & 1000 \\
 \hline
East Bay 		& 3125 & 3350 		& $d_g(\textrm{Bay Sensor}) = 3750$ & 	1500			& 575 & 2000 \\
 
 \hline
Airfield Office 	& 2100 & 2000 		& $d_g(\textrm{Sensor 3}) = 2375$ &	450			& 550 & 500 \\
 \hline
Airfield 		& 1750 & 2875 		& $d_g(\textrm{Sensor 3}) = 2375$ & 	700			& 950 & 1625 \\
\hline
\end{tabular}
\label{t:areaTaskDims}
\end{table*}
For an area search task $t_a \in \{$Bridge, Main Office, Warehouse, Truck Depot, Shipyard Office, Shipyard, West Bay, East Bay, Airfield Office, Airfield\}, 
let $d_g(t_a)$ be the distance from the Ground Control Station to the starting corner of the search, 
$d_g'(t_a)$ be the distance back to the Ground Control Station from the terminal corner of the search, 
and let $d_w(t_a)$ and $d_h(t_a)$ be the width and height of the area to be searched, 
as given in \tabref{t:lineTaskDims}.
In order to sweep an EO sensor footprint across the entire search area, 
a UAV has to make several parallel passes over the area along its height. 
Let $\bar{e}_w$ be the width of a UAV's EO sensor footprint at the UAV's standard operating altitude. 
Then the total distance $d(t_a)$ traveled to perform task $t_a$ is approximately
\begin{equation}
d(t_a) = d_g(t_a) + \frac{d_w(t_a)}{\bar{e}_w}d_h(t_a)+ d_g'(t_a).
\label{eq:areaDistances}
\end{equation}

When an area search is prompted by the detection of an intruder by a ground sensor, 
the distance calculation is modified to account for travel time to the sensor 
as well as a potentially different sensor footprint width $e_w$.
Let the total distance in response to a ground sensor be $d_s(t_a)$. 
Then 
\begin{equation}
d(t_a) = d_s(t_a) + d'_s(t_a) + \frac{d_w(t_a)}{e_w}d_h(t_a)+ d_g'(t_a).
\label{eq:sensorAreaDistances}
\end{equation}

\subsection{Image GSD}
\label{s:gsd}
\begin{figure}[thpb]
      \centering
      \scalebox{0.82}{\includegraphics[]{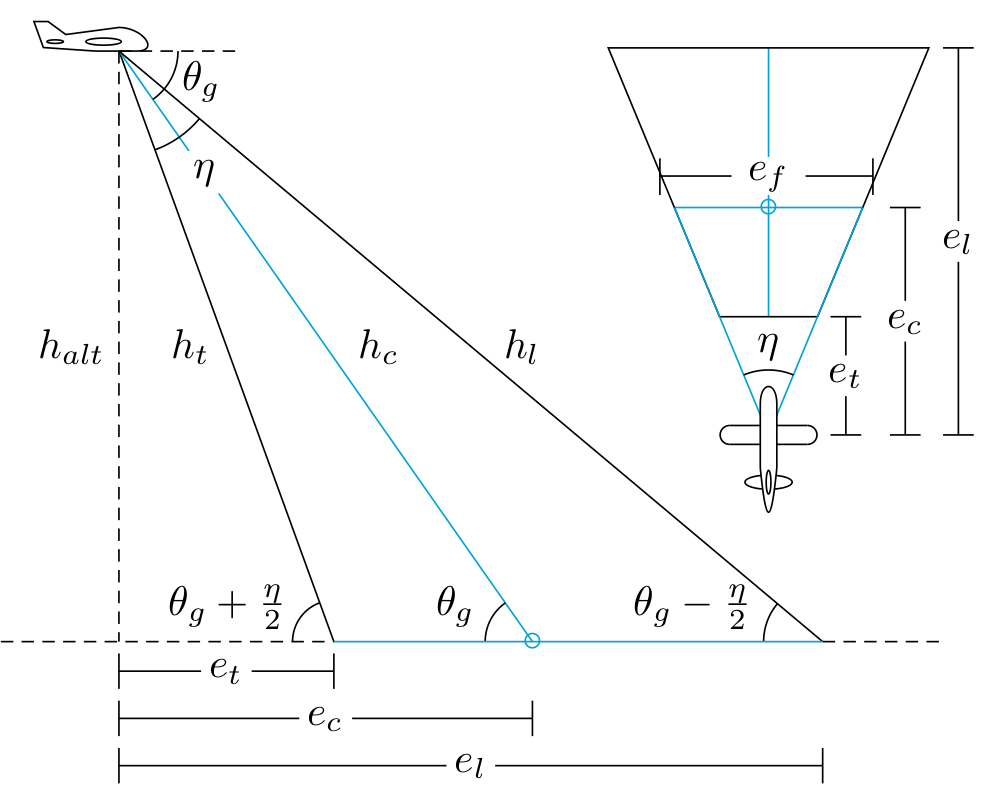}}
          \caption{The footprint of an EO sensor, which depends on gimbal elevation angle, field of view, and UAV altitude.}
      \label{fig:sensorFootprint1}
\end{figure}

We measure GSD $G$ as the number of pixels across the width of an EO sensor footprint at its center as shown in \figref{fig:sensorFootprint1}, resulting in the expression
\begin{equation}
G = \frac{e_w}{r_h} = \frac{2 h_{alt} \tan (\eta / 2) / \sin(\theta_g)}{r_h},
\label{e:gsdGeneral}
\end{equation}
where $e_w$ is the width of the EO sensor footprint and $r_h$ the horizontal resolution of the EO sensor.
Note that $e_w$ depends on UAV altitude $h_{alt}$, EO sensor field of view $\eta$, and EO sensor gimbal angle $\theta_g$. 

In the case study, most objects are large enough that GSD will not be a concern. However, intruders on foot near Sensor~1, Sensor 2, or Sensor 3, or intruders on small personal watercraft near the Bay Sensor Network will require a low GSD. For detecting intruders indicated by these ground sensors, we assume a minimum of $\pi/12$ for $\eta$, since smaller values tend to result in a shaky stream of video images due to noise in UAV flight trajectories, and we assume $\theta_g$ is set to $\pi/2$ so that the camera is pointing straight down. With $\eta = \pi/12$ and $\theta_g = \pi/2$, the expression for $G$ becomes
\begin{equation}
G = \frac{2 h_{alt} \tan(\pi/24)}{r_h}.
\label{e:gsd}
\end{equation}

\subsection{Object Recognition versus Altitude, Speed, and Sensor Quality}
\label{s:recognition}

The UAVs in the case study are essentially tools for carrying out various 
surveillance tasks in collaboration with human operators.
For each surveillance task, a UAV flies over the associated region and sends 
imagery back to the ground control station, where an operator analyzes it. 
The level of analysis the operator is able to perform depends on many factors, 
including the size of objects of interest relative to the resolution of the imagery  
and how long the operator is able to view objects in the imagery. 

Models of human performance for basic visual analysis tasks are often traced to research 
performed by Johnson in the context of night vision systems \cite{johnson_1958}. 
Johnson classified visual analysis tasks into several different categories, including
\emph{detection}, \emph{recognition}, and \emph{identification}. 
Detection is roughly defined as the ability to discern whether an object of significance is present in an image; 
recognition as determining the class of an object relative to similarly sized objects, e.g., a truck versus a car; 
and identification as distinguishing between specific members of the same class, e.g., a man versus a woman. 
Johnson characterized the probability of successfully performing these tasks as 
a function of the number of line pairs across the critical dimension of an object in an image,
where a line pair corresponds to two lines of pixels in digital imagery and 
critical dimension refers to the smaller dimension of an object. 
Much of this research was originally performed on targets with aspect ratios less than 2:1. 
Later work by Moser found that for objects with large aspect ratios of about 10:1, the number of line pairs or pixels 
across the area of the object is a better predictor of task performance \cite{moser_1972}. 
As given in \cite{driggers_1997}, Table \ref{table:n50criteria} lists $n_{50}$ values, i.e., numbers of line pairs 
across an object needed for a 50\% probability of successfully performing a visual analysis task on the object. 
Note that the 1D criterion applies to the critical dimension of an object with a ``small'' aspect ratio, 
while the 2D criterion applies to the geometric mean of the height and width of an object with a ``large'' aspect ratio.

\begin{table}[ht]
\caption{Johnson chart -- the number of line pairs needed for a 50\% probability of performing a visual analysis task on an object. }
\centering
\begin{tabular}{lccl}
\toprule
\textbf{Task type} & \textbf{1D} $n_{50}$ & \textbf{2D} $n_{50}$ & \textbf{Description} \\
\midrule
Detection & $1.0$ & 0.75 & Object is present \\
Recognition & $4.0$ & 3.0 & Class of object \\
Identification & $8.0$ & 6.0 & Class member of object  \\
\bottomrule
\end{tabular}
\label{table:n50criteria}
\end{table}

Suppose there are $n$ line pairs across an object in an image. 
Then given an unlimited amount of time to view the image, the probability of 
an operator being able to successfully perform a visual analysis task can be estimated as 
\begin{equation}
P_{\infty} = \frac{(n/n_{50})^{x_0}}{1+(n/n_{50})^{x_0}} \quad\textrm{where}\quad x_0 = 2.7 + 0.7(n/n_{50}).
\label{eq:pInf}
\end{equation}
Further details on \eqnref{eq:pInf} can be found in \cite{kopeika_1998}.

Suppose UAVs in the case study are equipped with standard electro-optical (EO) sensors.
Most EO sensors feature adjustable magnification, which allows the sensor's field of view to vary.
An EO sensor may also be mounted on a gimbal, so that it can be pointed in different directions relative to the UAV. 
As shown in \figref{fig:sensorFootprint1}, the size of the EO sensor's footprint on the ground 
depends on its gimbal elevation angle $\theta_g$, its vertical field of view $\eta_v$, 
its horizontal field of view $\eta_h$, and the altitude of the UAV $h_{alt}$. 
Suppose the vertical and horizontal fields of view are equal so that $\eta_h = \eta_v$. 
Suppose also that the sensor is pointed straight down so that $\theta_g = 90^\circ$. 
Then the height and width $d_w$ of the sensor footprint will be the same, with 
\begin{equation}
d_w = 2 h_{alt} \tan(\eta_h/2).
\label{eq:sensorWidth}
\end{equation}
The region that lies within the sensor footprint is projected onto a sensor array 
that converts light into electrical signals. 
The resolution of the resulting image depends on the resolution of this sensor array. 
As shown in Table \ref{table:eoCosts}, EO sensors with different resolutions can be purchased 
for different prices, where $r_v$ and $r_h$ are vertical and horizontal resolution in pixels.
Given the sensor resolution, current field of view, UAV altitude, and assuming $r_h > r_v$, we can  
calculate a conservative upper bound on the ground sample distance $gsd$ of the image in m/pixel as 
\begin{equation}
gsd = \frac{d_w}{r_v}.
\label{eq:sensorGsd}
\end{equation}
The number of line pairs $n$ across an object of size $d_o$ in the image is then 
\begin{equation}
n = \frac{d_o}{2 gsd} = \frac{d_o r_v } {2 d_w} = \frac{d_o r_v } {4 h_{alt} \tan(\eta_h/2)}.
\label{eq:objectLinePairs}
\end{equation}

\begin{table}[ht]
\caption{Hypothetical costs for EO sensors with different image resolutions.}
\centering
\begin{tabular}{cccc}
\toprule
 & \multicolumn{3}{c}{\textbf{Resolution} $r_h \times r_v$} \\ \cmidrule{2-4}
 & 640$\times$480 & 1280$\times$720 & 1920$\times$1080 \\
   & (480p) & (720p) & (1080p) \\
 \midrule
\textbf{Cost} & \$15k & \$30k & \$45k \\
\bottomrule
\end{tabular}
\label{table:eoCosts}
\end{table}

\subsection{UAV Altitude}
\label{s:altitude}

\begin{figure}[thpb]
      \centering
      \includegraphics[width=.65\columnwidth]{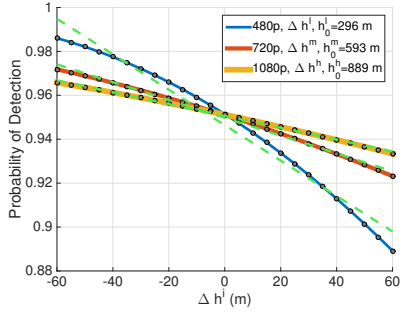}
      \caption{Probability of detecting an intruder for $\Delta h \in [-60, 60]$ for three EO sensor resolution options. Linear approximations are shown as dashed lines, and quadratic approximations are shown as traces of black dots.}
      \label{fig:gsdProbs}
\end{figure}

Since the probability that a human operator can analyze an object in an image defined in \eqref{eq:pJohnson} varies significantly with altitude $h_{alt}$, 
we define a different operating altitude for each type of EO sensor. 
Let $r_h^l$, $r_h^m$, and $r_h^h$ be the horizontal resolution of the 
low (480p), mid (720p), and high (1080p) resolution EO sensor options, respectively. 
Assume $d_o = 0.5$ m for intruders, both humans on foot and small watercraft. 
Then let us define operating altitudes $h_0^l$, $h_0^m$, and $h_0^h$ such that 
the probability of detection $p_d^l$, $p_d^m$, and $p_d^h$ for each EO sensor option is 
approximately 0.95 at the corresponding operating altitude. 
In this case, $h_0^l = 296$ m, $h_0^m = 593$ m, and $h_0^h = 889$ m. 
Suppose we allow altitude to vary by an amount $\Delta h^l, \Delta h^m, \Delta h^h \in [-60, 60]$ m around each operating altitude. 
Then \figref{fig:gsdProbs} shows the probability of detecting an intruder as $\Delta h^l, \Delta h^m,$ and $\Delta h^h$ 
vary for the three EO sensor options around operating altitudes $h_0^l$, $h_0^m$, and $h_0^h$.
Linear and quadratic approximations were computed using \textsc{Matlab}'s constrained linear least squares solver \emph{lsqlin} with the constraint that the approximated function have an upper bound of 1. Linear approximations, shown as dashed lines in \figref{fig:rocCurvesApprox}, are
\begin{alignat}{2}
p_{d}^l 	&= -0.0008 \Delta h^l + 0.9461 \quad && \text{for $\Delta h^l \in [-60, 60]$} \\
p_{d}^m  	&= -0.0004 \Delta h^m + 0.9498 \quad && \text{for $\Delta h^m \in [-60, 60]$} \\
p_{d}^h	&= -0.0003 \Delta h^h + 0.9505 \quad && \text{for $\Delta h^h \in [-60, 60]$},  
\end{alignat}
and quadratic approximations, plotted with dots in \figref{fig:rocCurvesApprox}, are
\begin{alignat}{2}
p_{d}^l 	&= -0.000004 {\Delta h^l}^2 - 0.000810 \Delta {h^l}+ 0.951075 \nonumber \\
& \quad \quad \text{for $\Delta h^l \in [-60, 60]$} \\
p_{d}^m  	&= -0.000001 {\Delta h^m}^2 - 0.0004051 \Delta {h^m} + 0.9511169 \nonumber \\
& \quad \quad \text{for $\Delta h^m \in [-60, 60]$} \\
p_{d}^h	&= -0.000000 {\Delta h^h}^2- 0.000300 \Delta {h^h} + 0.950500 \nonumber \\
& \quad \quad \text{for $\Delta h^h \in [-60, 60]$}, 
\end{alignat}
with linear and quadratic approximations for $p_{d}^h$ being equivalent.

\begin{figure}[thpb]
      \centering
      \includegraphics[width=.65\columnwidth]{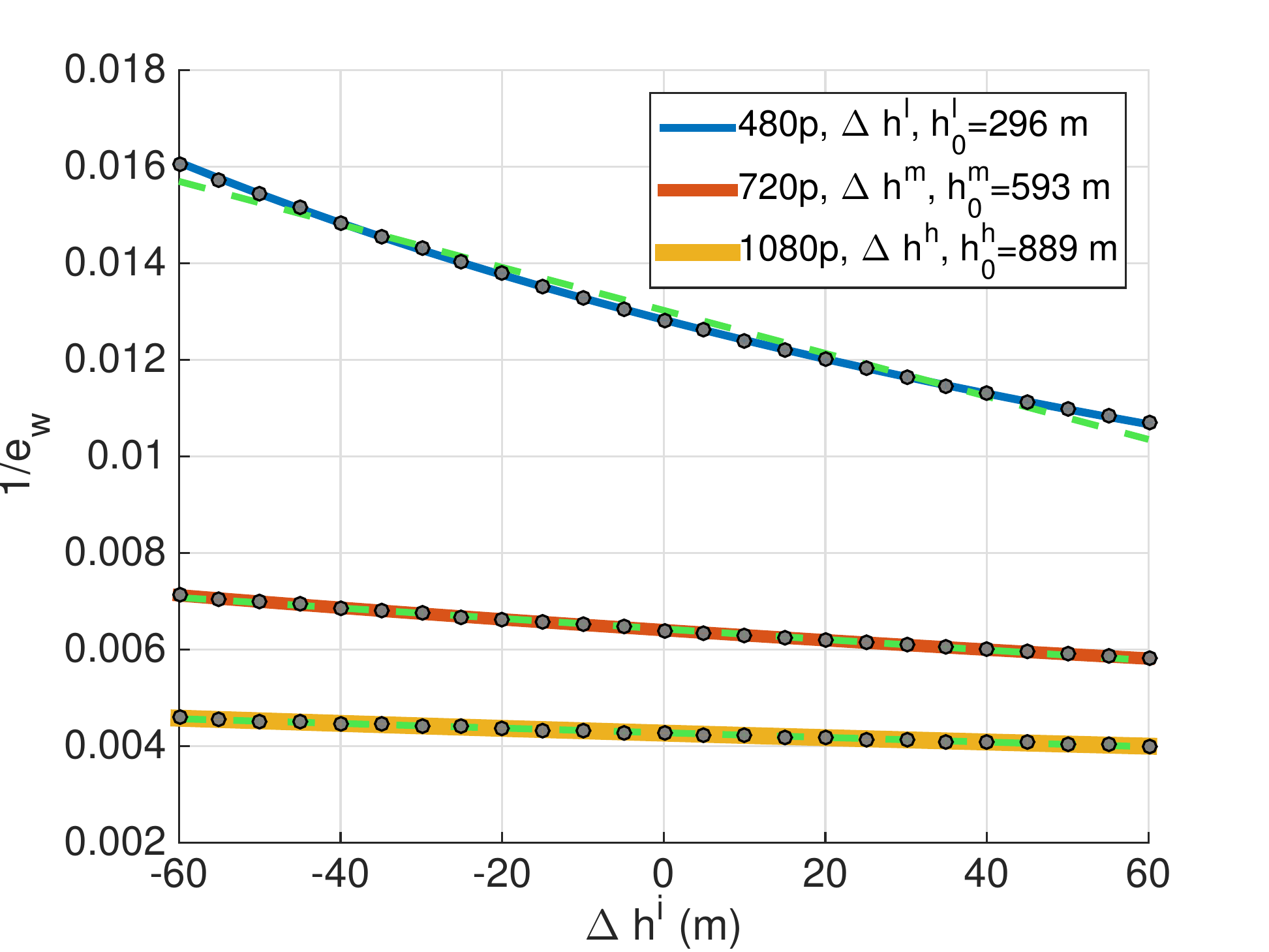}
      \caption{Plots of $1/e_w$ for $\Delta h \in [-60, 60]$ for three sensor resolution options. Linear approximations are shown as dashed lines, and quadratic approximations are shown as traces of black dots.}
      \label{fig:gsd}
\end{figure}

The altitude at which a UAV performs an area search will also change the amount of time needed to perform the search. 
This is because the width of the EO sensor footprint $e_w$ changes with altitude, 
and in order to sweep the EO sensor footprint over the entire area, 
a UAV must make approximately $d_w/e_w$ passes of length $d_h$ over the area, 
where $d_w$ is the task area width and $d_h$ task area height, as given in \tabref{t:areaTaskDims}.
Let us consider how this changes the operational cost of an area search when searching for intruders. 
Suppose cost per flight hour is approximately 1,000 dollars per hour or $c_f = 5/18$ dollars per second.
Suppose a UAV's standard operating altitude $h_0$ for surveillance tasks other than intruder area searches is $h_0 = h_0^h$, 
its standard ground speed $v_g$ is $15$ m/s, 
and its speed $v_a$ to ascend or descend during altitude changes is $5$ m/s. 
Suppose for intruder area searches, a UAV changes altitude from $h_0$ to $h_0^i + \Delta h^i$ 
for $\Delta h^i \in [-60,60]$ m, $i \in \{l, m, h\}$ before a search, then changes back to $h_0$ after a search. 
Then the modified cost for an area search task $t_a$ for intruders is approximately
\begin{align}
c(t_a) =  \frac{5}{18} \Big(& d_s(t_a) + d'_s(t_a) + \frac{1}{e_w}\frac{d_w(t_a) d_h(t_a)}{v_g} + \mbox{} \nonumber \\
& \frac{2 (h_0 - h_0^i - \Delta h^i)}{v_a} +  d'_g(t_a)\Big),
\label{eq:baseCost}
\end{align}
which replaces the expression for cost \eqnref{e:basicCost} given in Section \ref{sec:basicTaskCosts}.
\figref{fig:gsd} shows how $1/e_w$ changes for the low, mid, and high cost EO sensor options given their operating points.
In our approach, we obtain the following linear approximations to use in \eqnref{eq:baseCost}, shown as dashed lines in \figref{fig:gsd}, 
\begin{alignat}{2}
1/e_w^l	&= -0.000045 \Delta h^l + 0.013026 \quad && \text{for $\Delta h^l \in [-60, 60]$} \\
1/e_w^m 	&= -0.000011 \Delta h^m + 0.006428 \quad && \text{for $\Delta h^m \in [-60, 60]$} \\
1/e_w^h	&= -0.000005 \Delta h^h + 0.004279 \quad && \text{for $\Delta h^h \in [-60, 60]$},
\end{alignat}
and the following quadratic approximations to use in  \eqnref{eq:baseCost}, plotted as black dots in \figref{fig:gsd},  
{\small
\begin{alignat}{2}
1/e_w^l	&= 0.0000002 {\Delta h^l}^2 - 0.0000445 h^l + 0.0128284 \nonumber \\ 
& \quad \quad \text{for $\Delta h^l \in [-60, 60]$} \\
1/e_w^m  	&= -0.0000000 {\Delta h^m}^2 - 0.0000110 h^m + 0.0064280 \nonumber \\
& \quad \quad \text{for $\Delta h^m \in [-60, 60]$} \\
1/e_w^h	&= -0.0000000 {\Delta h^h}^2 - 0.0000050 h^h + 0.0042790 \nonumber \\
& \quad \quad \text{for $\Delta h^h \in [-60, 60]$,}
\end{alignat}
}%
with $1/e_w^m$ and $1/e_w^h$ each having equivalent linear and quadratic approximations.

%
%

\subsection{ROC Curve Approximations}
\label{s:roc}

Approximations of the ROC curves were computed using \textsc{Matlab}'s constrained linear least squares solver \emph{lsqlin} with the constraint that the approximated function have an upper bound of 1. Let $p_{f}^l$, $p_{f}^m$, and $p_{f}^h$ be the false positive rate for low, mid, and high cost sensors respectively, and let $p_{t}^l$, $p_{t}^m$, and $p_{t}^h$ be the true positive rate for low, mid, and high cost sensors respectively. Then we have the following linear approximations, shown as dashed lines in \figref{fig:rocCurvesApprox},
\begin{alignat}{2}
p_{t}^l 	&= 0.0041 p_{f}^l + 0.9949 \quad && \text{for $p_{f}^l \in [0.2, 1.0]$} \\
p_{t}^m  	&= 0.0853 p_{f}^m + 0.9137 \quad && \text{for $p_{f}^m \in [0.2, 1.0]$} \\
p_{t}^h	&=  \quad && \text{for $p_{f}^h \in [0.2, 1.0]$},  
\end{alignat}
and the following quadratic approximations, plotted with dots in \figref{fig:rocCurvesApprox},
\begin{alignat}{2}
p_{t}^l 	&= -0.0183 {p_{f}^l}^2 + 0.0273 p_{f}^l + 0.9888 \quad && \text{for $p_{f}^l \in [0.2, 1.0]$} \\
p_{t}^m  	&= -0.2243 {p_{f}^m}^2 + 0.3779 p_{f}^m + 0.8391 \quad && \text{for $p_{f}^m \in [0.2, 1.0]$} \\
p_{t}^h	&= -0.4676 {p_{f}^h}^2 + 0.9213 p_{f}^h + 0.5346 \quad && \text{for $p_{f}^h \in [0.2, 1.0]$}. 
\end{alignat}
Let $q_f$ be the false negative rate and $q_t$ be the true negative rate. Then note that given a true positive rate $p_t$ and a false positive rate $p_f$, $q_f = 1 - p_t$ and $q_t = 1 - p_f$.

%% file: main.bbl
\begin{thebibliography}{10}

\bibitem{johnson_1958}
J.~Johnson.
\newblock Analysis of image forming systems.
\newblock In {\em Image Intensifer Symposium}, pages 249 -- 273, 1958.

\bibitem{moser_1972}
P.~M. Moser.
\newblock Mathematical model of {FLIR} performance.
\newblock Technical Report Technical Memorandum NADC-20203, Naval Air
  Development Center, Warminster, PA, 1972.

\bibitem{DBLP:journals/ior/SatiaL73}
Jay~K. Satia and Roy E.~Lave Jr.
\newblock Markovian decision processes with uncertain transition probabilities.
\newblock {\em Operations Research}, 21(3):728--740, 1973.

\bibitem{driggers_1997}
R.~Driggers, M.~Kelley, and Paul Cox.
\newblock National imagery interpretation rating system and the probabilities
  of detection, recognition, and identification.
\newblock {\em Optical Engineering}, 36(7):1952--1959, 1997.

\bibitem{RR98}
Michael~K. Reiter and Aviel~D. Rubin.
\newblock Crowds: Anonymity for web transactions.
\newblock {\em ACM Trans.\ on Information and System Security}, 1(1):66--92,
  1998.

\bibitem{Las01}
Jean~B. Lasserre.
\newblock Global optimization with polynomials and the problem of moments.
\newblock {\em {SIAM} Journal on Optimization}, 11(3):796--817, 2001.

\bibitem{Zeroconf03}
Henrik Bohnenkamp, Peter Van~Der Stok, Holger Hermanns, and Frits Vaandrager.
\newblock Cost-optimization of the {IPv4} zeroconf protocol.
\newblock In {\em DSN}, pages 531--540. IEEE CS, 2003.

\bibitem{daws04}
Conrado Daws.
\newblock Symbolic and parametric model checking of discrete-time {M}arkov
  chains.
\newblock In {\em ICTAC}, volume 3407 of {\em LNCS}, pages 280--294. Springer,
  2004.

\bibitem{fawcett_2006}
Tom Fawcett.
\newblock An introduction to {ROC} analysis.
\newblock {\em Pattern Recognition Letters}, 27(8):861--874, 2006.

\bibitem{DBLP:journals/lmcs/EtessamiKVY08}
Kousha Etessami, Marta~Z. Kwiatkowska, Moshe~Y. Vardi, and Mihalis Yannakakis.
\newblock Multi-objective model checking of {M}arkov decision processes.
\newblock {\em Logical Methods in Computer Science}, 4(4), 2008.

\bibitem{param_sttt}
Ernst~Moritz Hahn, Holger Hermanns, and Lijun Zhang.
\newblock Probabilistic reachability for parametric {M}arkov models.
\newblock {\em Software Tools for Technology Transfer}, 13(1):3--19, 2010.

\bibitem{KNP11}
Marta Kwiatkowska, Gethin Norman, and David Parker.
\newblock \textsc{Prism} 4.0: Verification of probabilistic real-time systems.
\newblock In {\em CAV}, volume 6806 of {\em LNCS}, pages 585--591. Springer,
  2011.

\bibitem{DBLP:conf/atva/ForejtKP12}
Vojtech Forejt, Marta~Z. Kwiatkowska, and David Parker.
\newblock Pareto curves for probabilistic model checking.
\newblock In {\em ATVA}, volume 7561 of {\em LNCS}, pages 317--332. Springer,
  2012.

\bibitem{KNP12b}
Marta Kwiatkowska, Gethin Norman, and David Parker.
\newblock The {PRISM} benchmark suite.
\newblock In {\em QEST}, pages 203--204. IEEE CS, 2012.

\bibitem{DBLP:conf/tacas/ChenFKPS13}
Taolue Chen, Vojtech Forejt, Marta~Z. Kwiatkowska, David Parker, and Aistis
  Simaitis.
\newblock Prism-games: {A} model checker for stochastic multi-player games.
\newblock In {\em TACAS}, volume 7795 of {\em LNCS}, pages 185--191. Springer,
  2013.

\bibitem{DBLP:conf/csl/BaierDK14}
Christel Baier, Clemens Dubslaff, and Sascha Kl{\"{u}}ppelholz.
\newblock Trade-off analysis meets probabilistic model checking.
\newblock In {\em CSL-LICS}, pages 1:1--1:10. {ACM}, 2014.

\bibitem{casbeer_2014}
Hua Chen, Krishna Kalyanam, Wei Zhang, and David Casbeer.
\newblock Continuous-time intruder isolation using {U}nattended {G}round
  {S}ensors on graphsround sensors on graphs.
\newblock In {\em ACC}, 2014.

\bibitem{DBLP:conf/icse/GordonHNR14}
Andrew~D. Gordon, Thomas~A. Henzinger, Aditya~V. Nori, and Sriram~K. Rajamani.
\newblock Probabilistic programming.
\newblock In {\em FoSER}, pages 167--181. ACM Press, 2014.

\bibitem{iscasmc}
Ernst~Moritz Hahn, Yi~Li, Sven Schewe, Andrea Turrini, and Lijun Zhang.
\newblock {iscasMc}: {A} web-based probabilistic model checker.
\newblock In {\em FM}, volume 8442 of {\em LNCS}, pages 312--317. Springer,
  2014.

\bibitem{dehnert-et-al-cav-2015}
Christian Dehnert, Sebastian Junges, Nils Jansen, Florian Corzilius, Matthias
  Volk, Harold Bruintjes, Joost{-}Pieter Katoen, and Erika {\'{A}}brah{\'{a}}m.
\newblock Prophesy: {A} probabilistic parameter synthesis tool.
\newblock In {\em {CAV} {(1)}}, volume 9206 of {\em LNCS}, pages 214--231.
  Springer, 2015.

\bibitem{garcia2015comprehensive}
Javier Garc{\i}a and Fernando Fern{\'a}ndez.
\newblock A comprehensive survey on safe reinforcement learning.
\newblock {\em Journal of Machine Learning Research}, 16(1):1437--1480, 2015.

\bibitem{param_website}
{\tool{PARAM} Website}, 2015.
\newblock \url{http://depend.cs.uni-sb.de/tools/param/}.

\bibitem{kingston_2015}
Steven Rasmussen and Derek Kingston.
\newblock Development and flight test of an area monitoring system using
  {U}nmanned {A}erial {V}ehicles and {U}nattended {G}round {S}ensors.
\newblock In {\em International Conference on Unmanned Aircraft Systems}, pages
  1215--1224, 2015.

\bibitem{DBLP:journals/ai/DelgadoBDS16}
Karina~Valdivia Delgado, Leliane~N. de~Barros, Daniel~B. Dias, and Scott
  Sanner.
\newblock Real-time dynamic programming for markov decision processes with
  imprecise probabilities.
\newblock {\em Artif. Intell.}, 230:192--223, 2016.

\bibitem{junges-et-al-tacas-2016}
Sebastian Junges, Nils Jansen, Christian Dehnert, Ufuk Topcu, and
  Joost{-}Pieter Katoen.
\newblock Safety-constrained reinforcement learning for mdps.
\newblock In {\em TACAS}, volume 9636 of {\em LNCS}, pages 130--146. Springer,
  2016.

\bibitem{kingston2016automated}
Derek Kingston, Steven Rasmussen, and Laura Humphrey.
\newblock Automated {UAV} tasks for search and surveillance.
\newblock In {\em CCA}, pages 1--8. IEEE, 2016.

\bibitem{quatmann-et-al-atva-2016}
Tim Quatmann, Christian Dehnert, Nils Jansen, Sebastian Junges, and
  Joost{-}Pieter Katoen.
\newblock Parameter synthesis for markov models: Faster than ever.
\newblock In {\em {ATVA}}, volume 9938 of {\em LNCS}, pages 50--67, 2016.

\bibitem{storm}
Christian Dehnert, Sebastian Junges, Joost{-}Pieter Katoen, and Matthias Volk.
\newblock A storm is coming: {A} modern probabilistic model checker.
\newblock In {\em CAV}, volume 10427 of {\em LNCS}, pages 592--600. Springer,
  2017.

\bibitem{profeat}
Philipp Chrszon, Clemens Dubslaff, Sascha Kl{\"{u}}ppelholz, and Christel
  Baier.
\newblock Profeat: feature-oriented engineering for family-based probabilistic
  model checking.
\newblock {\em Formal Asp. Comput.}, 30(1):45--75, 2018.

\bibitem{arnd_tacas}
Arnd Hartmanns, Sebastian Junges, Joost-Pieter Katoen, and Tim Quatmann.
\newblock Multi-cost bounded reachability in mdps.
\newblock In {\em TACAS}, LNCS, April 2018.

\bibitem{BK08}
Christel Baier and Joost-Pieter Katoen.
\newblock {\em Principles of Model Checking}.
\newblock The MIT Press, 2008.

\bibitem{bertsekas1999nonlinear}
Dimitri~P. Bertsekas.
\newblock {\em Nonlinear Programming}.
\newblock Athena Scientific Belmont, 1999.

\bibitem{kopeika_1998}
N.~S. Kopeika.
\newblock {\em A System Engineering Approach to Imaging}.
\newblock SPIE Press, 1998.

\bibitem{sut98a}
R.S. Sutton and A.G. Barto.
\newblock {\em Reinforcement Learning -- An Introduction}.
\newblock MIT Press, 1998.

\end{thebibliography}
